\documentclass[aps,nofootinbib,showpacs,preprintnumbers,amsmath,amssymb,floatfix]{revtex4}
\def\be{\begin{equation}}
\def\ee{\end{equation}}
\def\be*{\begin{equation*}}
\def\ee*{\end{equation*}}
\def\bea{\begin{eqnarray}}
\def\eea{\end{eqnarray}}
\def\bea*{\begin{eqnarray*}}
\def\eea*{\end{eqnarray*}}

\usepackage{graphics}
\usepackage{graphicx}
\usepackage{dcolumn}
\usepackage{bm}
\usepackage{epsfig}
\usepackage{graphicx}
\usepackage{multirow}

\begin{document}

\vspace{1cm}

\preprint{hep-ph/9208225v2, published in Int.~J.~Mod.~Phys. A9, 5313 (1994)}

\title{Two- and Three Vector Boson Production in ${\bf e^+e^-}$ Collisions within the BESS Model}

\author{G.~Cveti\v{c}\footnote{Address since Sept.~2000: Dept.~of Physics, Univ. T\'ecnica Federico Santa Mar\'{\i}a, Valpara\'{\i}so, Chile}}
\affiliation{Institut f\"ur Physik, Universit\"at Dortmund, W-4600 Dortmund 50}

\author{C.~Grosse-Knetter, and R. K\"ogerler}
\affiliation{Fakult\"at f\"ur Physik, Universit\"at Bielefeld,
33501 Bielefeld, Germany}

\begin{abstract}
The BESS model is the Higgsless alternative to the standard model of
electroweak interaction with nonlinear realized spontaneous symmetry
breaking. Since it is non-renormalizable new couplings (not existing
in SM) are induced at each loop order. On the basis of the one-loop 
induced gauge boson self-couplings we calculated the cross
sections for the two- and three-gauge-boson production processes in
$e^+e^-$ collisions. Measurements of these cross sections in a planned
$e^+e^-$ linear collider at $\sqrt s = 500 \ {\rm GeV}$ (NLC) will supply a
good empirical test of the gauge boson self-interactions and thus
should enable to discriminate between SM and the BESS model.
\end{abstract}

\maketitle  

\section{Introduction}
A thorough experimental
investigation of the gauge boson self-interactions is of
utmost importance for identification of the ``true'' (gauge-) theory
of electroweak interactions. The most powerful instrument for such
an analysis is certainly provided by production processes of (two
and three) gauge bosons in electron-positron annihilation at
sufficiently high energy. For the conclusiveness
of such experiments the available energy plays an important role,
since
most alternatives to SM of electroweak interactions
are characterized by deviations from the Yang
Mills type self-couplings. Hence they lead to deviations from SM
predictions which in general increase with increasing energy.
Therefore, although LEP II is at the horizon and will certainly yield
interesting results, the efforts for establishing an
$e^+e^-$-collider at energies far beyond the $W^+W^-$ threshold (NLC)
\cite{Proc1} go into the right directions.
It is our conviction, and we will give indications within the present
paper, that new physics can be tested with sufficient reliability if
the energy of such a machine lies beyond $350 \ {\rm GeV}$.\par
There are several possible ways for testing gauge boson self
interactions in boson production. One possibility starts from the
most general interaction Lagrangian (both for three and four boson
vertices) being expressed in terms of unspecified coupling constants.
By investigating sufficiently many physical quantities (cross sections
with different polarization configurations, asymmetries, density
matrices etc.) it should be possible in principle to determine those
coupling strengths numerically.
Due to the complexity of the general Lagrangians,
however, and because of possible conspiracies between different
terms, such a model-independent analysis will hardly be feasible
in practice \cite{Prep2}. Consequently, a realistic analysis has to be footed on
specific models, i.e., attainable experimental results are to
be interpreted within given models. In this way it should be possible
to discriminate between various theoretical possibilities for the
vector boson self-interactions and, in particular, decide how well
SM of electroweak interaction is verified by experimental data.

Within the present and a forthcoming paper we are applying the latter
procedure for investigating vector boson self couplings within the
so-called BESS model 
\cite{UGVA-DPT-1986-01-492,BI-TP-88/32}. 
This is considered to be one of the most
attractive alternatives to SM, since it does not represent a
trivial extension of the latter one (by simply adding further gauge
groups together with the appropriate Higgs fields), but is footed
on a different mechanism for gauge boson mass generation
which completely avoids physical scalar (Higgs) particles. In a
more general sense it represents the most economical way of
parametrizing the effects of a strongly interacting sector of
(longitudinally polarized) gauge bosons.

Since any Higgs-less field theory of massive vector bosons is
non-renormaliz\-able and has to be considered as an effective
theory, it is of utmost importance to clarify in detail the
quantum effects (emerging from loop-generated interactions)
contributing to the different reactions.
A thorough analysis of these quantum induced interactions has been
completed recently 
\cite{BI-TP-88/32,BI-TP-90-43} and the results will be converted into
predictions for various physical (boson production) processes within
the present paper. First results of this analysis have already
been published \cite{it6}. Here, we present the full wealth of BESS-model
predictions (to one-loop order) for those reactions which will be
feasible with $e^+e^-$-colliders working in an energy range of
above 350 GeV. In another paper \cite{hep-ph/9212291} 
will be devoted to utilizing
the present results for a careful identification of the allowed
(BESS-model-) parameter range as restricted by the accuracy
which can be reached in such experiments.\par
In Sec.~II we will motivate the BESS model and roughly sketch its
main features. Section III is devoted to a general discussion of its
phenomenological structure, mainly with respect to experiments at
high energy $e^+e^-$-colliders. Section IV contains
the presentation and
discussion of our results. In Sec.~V we draw some final conclusions.

\section{The BESS-Model}
We refrain from presenting the details of the model, since it has
been described already several times 
\cite{UGVA-DPT-1986-01-492,BI-TP-88/32}, and will only
sketch the theoretical motivations
and describe its main features concerning phenomenology.\par
The model can be motivated in several ways. Probably the most
fundamental and simple one starts from the sheer existence of massive
vector bosons $(W^\pm, Z)$ and continues with the following line of
reasoning: It is a basic fact of field theory that massive vector
bosons (with Yang-Mills-type self coupling) can -- by a suitable
field-enlarging point transformation -- be embedded in a theory with
local gauge symmetry \cite{104360}. The standard version of this
transformation is the St\"uckelberg formalism \cite{879856} based on
(unconstrained) vector and (unphysical) scalar fields. The local
gauge symmetry cannot, however, be completely Wigner-realized (in
the St\"uckelberg picture, e.g., the scalars transform \`a la
Goldstone-Nambu), i.e., this symmetry is necessarily spontaneously
broken. The canonical version of this
spontaneous symmetry breaking (SSB) is the Higgs mechanism
as applied in the Weinberg-Salam
model and in all its trivially extended versions.
If one prefers to avoid physical scalar (Higgs) particles -- as we
do here -- the only way is by assuming that the
symmetry (breaking) is realized nonlinearly (NL) (i.e., the scalar
components of the spin-1-field operator transforms nonlinearly under
local gauge transformations and, consequently, can be completely
gauged away). When this mechanism is formally applied to the
$SU(2) \times U(1)$ symmetry of SM
one obtains the gauged version of the
nonlinear $\sigma$-model \cite{YTP-80-01}. 
Now it can be shown on the same field
theoretical footing as described before (field-enlarging point
transformation) that this theory is gauge-equivalent to theories
with additional (``hidden'') local symmetry groups 
\cite{RRK 84-22}. Again,
these symmetries can become formally apparent when appropriate
numbers of unphysical scalar (would be Goldstone) fields are
introduced into the Lagrangian. In general, the gauge bosons
connected to the additional local gauge groups could, in principle,
be interpreted as purely auxiliary fields (combinations of
unphysical scalars) with no direct physical consequences -- in
accordance with the fact that they are ``produced'' by point
transformations. However, if the starting theory is a nonlinear
one there are strong indications \cite{RRK 84-22} that these (a priori
hidden) gauge bosons will show up as physical particles. What has
been ``shown'' in fact 
\cite{ITEP-62-1978}, is the quantum-generation of kinetic terms
for these vector bosons which enables them to propagate as real
physical particles.\footnote{For two- and three-dimensional models, this
result has been generally proven; for four-dimensional models
it has only been derived at the one-loop level \cite{ITEP-62-1978}.}

In the case of a starting (nonlinearly realized) $SU(2) \times U(1)$
gauge symmetry, the additional (``hidden'') gauge groups are bound
to be of SU(2)-type \cite{CERN-TH-4876/87}. 
In the most simple case (which should
correctly describe physics at moderately low energies (up to 3 TeV,
as we will see later)) we thus have as a local gauge group
$SU(2)_L \times U(1)_Y \times SU(2)_V$ with the corresponding
gauge bosons $\vec {\tilde W}_\mu, \tilde Y_\mu, \vec {\tilde V}_\mu$
but with only 6 unphysical scalars (denoted by $\vec \pi$ and
$\vec \sigma$) and no physical Higgs bosons. The corresponding
(tree-level-) Lagrangian defines the BESS model 
\cite{UGVA-DPT-1986-01-492,BI-TP-88/32}.
There are five fundamental parameters: $g, g^\prime,
g^{\prime\prime}$ (the gauge coupling constants connected with the
three fundamental gauge groups $SU(2)_L, U(1)_Y, SU(2)_V$,
respectively), $f^2$ (the overall scale parameter measuring the
size of SSB) and $\lambda^2$ (the relative strength of the additional
``hidden'' symmetry).
An important part of the BESS model Lagrangian are the couplings of
(unphysical) scalars to the vector bosons. They provide both mass-
and mixing-terms of bosons. Consequently, the physical
(mass-eigenstate) vector-boson-fields (denoted by $W^\pm, Z, A, V^\pm,
V^0$) are mixtures of the original (unmixed) fields $\vec {\tilde W},
\tilde Y, \vec {\tilde V}$, the connection being given by two
mixing matrices (for charged and neutral particles separately):
$$
{g \tilde W^\pm \choose {{g^{\prime\prime}} \over 2}\tilde V^\pm} =
{\cal C} {W^\pm
\choose  V^\pm}~~~~~~~{\rm and}~~~~~~~
\left( 
\begin{matrix} g^\prime \tilde Y \cr
g \tilde W_3 \cr
{{g^{\prime\prime}} \over 2} \tilde V_3\cr 
\end{matrix}
\right) =
{\cal N}
\left( 
\begin{matrix} A \cr
Z \cr
V^0\cr 
\end{matrix} 
\right) \ , \eqno(2.1 a,b) 
$$
with
$$
{\cal C} = \left( 
\begin{matrix} 
g \cos \varphi & g \sin \varphi \cr
- {{g^{\prime\prime}} \over 2}
\sin \varphi & {{g^{\prime\prime}}\over 2} \cos
\varphi\cr
\end{matrix}
\right) \ , \eqno(2.2)
$$
and
$$
{\cal N} = {{gg^\prime} \over G} 
\left( 
\begin{matrix}
\cos \psi &(\sin \xi \sin \psi - {{g^\prime} \over g} \cos \xi)
&(- \cos \xi \sin \psi - {{g^\prime} \over g} \sin \xi) \cr
\cos \psi & (\sin \xi \sin \psi + {g \over {g^\prime}} \cos \xi)
& (- \cos \xi \sin \psi + {g \over {g^\prime}} \sin \xi) \cr
\cos \psi &- \sin \xi \sin \psi \cot^2 \psi &\cos \xi \sin \psi
\cot^2 \psi \cr
\end{matrix}
\right)~~.\eqno(2.3)
$$
The masses and mixing angles can be expressed in terms of the
model's parameters in the following compact way:\par
First define three quantities $p_i, c_i, d_i$ (for both the charged
$(i=1)$ and neutral $(i=2)$ sector) by
\bea*
p_1 &= 2 \lambda gg^{\prime\prime} \ , \qquad\qquad & p_2 = 2 {\tilde \lambda}
GG^{\prime\prime} \ , 
\qquad\qquad
\;\;\;\;\;\;\;\;\;\;\;\;\;\;\;\;\;\;\;\;\;\;\;\;\;\;\;\;\;\;\;\; (2.4) \cr
\\
c_1 &= g^2 (1 + \lambda^2) \ ,
\qquad\qquad & c_2 = G^2 (1 + {\tilde \lambda}^2) \ ,
\qquad\qquad
\;\;\;\;\;\;\;\;\;\;\;\;\;\;\;\;\;\;\;\;\;\;\;\;\;\;\; (2.5)\cr
\\
d_1 &=  g^{\prime\prime 2} \lambda^2 \ ,
\qquad\qquad & d_2 = G^{\prime\prime 2} {\tilde \lambda}^2 \ ,   
\qquad\qquad
 \;\;\;\;\;\;\;\;\;\;\;\;\;\;\;\;\;\;\;\;\;\;\;\;\;\;\;\;\;\;\;\;\;\; (2.6)\cr 
\eea*
(with ${\tilde \lambda} = \lambda (g^2 - g^{\prime 2})/(g^2 + g^{\prime 2}),
~~~G^2 = g^2 + g^{\prime 2},~~~G^{\prime\prime
2} = (\lambda^2/{\tilde \lambda}^2) (g^{\prime\prime 2} + 4 g^2 g^{\prime 2}/G^2)$) and
$$
e_i = c_i + d_i~~,~~~~~x_i = \sqrt{1- ({{p_i} \over {e_i}})^2}~~~~~
(i = 1,2).\eqno(2.7)
$$
Then the masses are given by
\bea*
M^2_{W^\pm} = \frac{1}{8} f^2  e_1 (1-x_1) \ , ~~~~~~~
& M^2_Z =  \frac{1}{8} f^2 e_2 (1-x_2) \ , 
\qquad\qquad\qquad\qquad &(2.8a,b)\cr
M^2_{V^\pm} = \frac{1}{8} f^2   e_1 (1+x_1) \ , ~~~~~~~
& M^2_{V^0} = \frac{1}{8} f^2  e_2 (1 + x_2) \ ,
\qquad\qquad\qquad\qquad &(2.9a,b)\cr
\eea*
and the mixing angles by
\bea*
tg (2 \varphi) &=& \lambda \frac{p_1}{(c_1 - d_1)} \ , \qquad
tg (2 \xi) = {\tilde \lambda} \frac{p_2}{(c_2 - d_2)}  \ , \qquad  \; 
\qquad\qquad\qquad\qquad\qquad (2.10a,b)
\nonumber\\
tg \psi &=& \frac{2gg^\prime}{G} \cdot \frac{1}{g^{\prime\prime}} \ .
\qquad \;\; 
\qquad\qquad\qquad\qquad\qquad (2.10c) 
\eea*
Note that the masses of the heavy bosons can be expressed in terms
of the corresponding light boson masses by means of the master
formula
$$
M_{V_i} = M_{W_i} {{1 + x_i} \over {1 - x_i}}~~~~~i = 1 ({\rm charged})~
{\rm or}~2 ({\rm neutral})~.\eqno(2.11)
$$
It allows an easy understanding of how the theory behaves as
$M_{V_i} \to \infty$. This limit is reached when $x_i \to 1$ or,
equivalently $(p_i/e_i) \to 0$. The latter condition can
be realized in three alternative ways:
\begin{itemize}
\item{(i)} $p_i \to 0$, i.e., $\lambda \to 0$ ($g, g^\prime$ finite
$\not= 0$, $g^{\prime\prime}$ finite or $0$)\footnote
{Note that
the electromagnetic coupling constant
${\scriptstyle e = \sqrt{4 \pi \alpha} =
(g g^\prime}/G) \cos \psi$, therefore neither ${\scriptstyle g}$
nor
${\scriptstyle g^\prime}$ nor
${\scriptstyle g^{\prime\prime}}$ can be equal to zero.}\hfill\break
In that case
$\varphi \to 0, \xi \to 0$; $\psi$ remains finite, but $V^{\pm 0}$
decouple from fermionic interactions.
\item{(ii)} $e_i \to \infty$ implied by
$g^{\prime\prime} \to \infty ~(\lambda^2 \not= 0)$\hfill\break
Here all mixing
angles vanish and we obtain SM ($V's$ decouple completely).
\item{(iii)} $e_i \to \infty$ implied by $\lambda^2 \to \infty$
($g^{\prime\prime}$ finite)\hfill\break
In this case all mixing angles are nonzero, i.e., the existence of
the (infinitely heavy) $V's$ manifest
themselves in the fermionic couplings. Furthermore,
as will be seen, the induced interactions increase with increasing
values of $M_V$. So, in this limit, the $V$-bosons do not decouple from
the low energy physics.\par
\end{itemize}
The effect of non-decoupling can be easily understood also by
remembering that for sufficiently heavy vector mass particles
$V^{\pm, 0}$ their masses are approximately given by
$$
M_{V^0} \simeq M_{V^\pm} \simeq M_W ({{g^{\prime\prime}} \over g})
\lambda \ , \eqno(2.12)
$$
i.e., the V-masses are driven by the (dimensionless) coupling constant
$\lambda$. This non-decoupling nature of V will be of importance in
particular for quantum-induced interactions as will be shown in the
following.\par
Phenomenological reasons [low energy
experiments are well reproduced by $SU(2)_L\break \times U(1)_Y$ alone]
imply that $V^\pm, V^0$ are heavy and the coupling constant $g^{\prime\prime}$
is large \cite{BI-TP-90-40}, since then the masses and the couplings of the light
particles are only slightly affected.

The couplings of fermions to vector bosons are specified by their
transformation property under the fundamental symmetry group, in
particular by the fact that they are singlets under the ``hidden''
group $SU(2)_V$. Thus fermions are coupled -- in the tree level
Lagrangian -- only to $\vec {\tilde W}_\mu$ and $\tilde Y_\mu$,
i.e., they interact with (physical) $V^{\pm, 0}$ only via mixing.
Similarly, the tree level structure of vector boson self interactions
is determined by the fact that the unmixed bosons $\vec {\tilde V}$
couple only among themselves (in Yang-Mills-manner) and thus
interactions between light and heavy (physical) bosons are mediated
by mixing only.\par
A particularly momentuous feature of the BESS model, which is
directly connected to its NL nature and to the consequent absence
of physical Higgs particles, is its lack of formal renormalizability,
which implies that the theory has to be understood as an effective
one, with finite validity range defined by a cut-off $\Lambda$.
As a consequence, cut-off dependent terms arise at 1- or
higher-loop level. They can be divided into two families: One group
consists of terms which can be fully absorbed into the starting
(tree-level) Lagrangian by renormalizing its field and/or coupling
constants. The remaining ones (which cannot be absorbed by
renormalization) constitute new observable interactions, which have
to be taken into account necessarily 
\cite{YTP-80-01} if the BESS model is taken
seriously. These contributions have been calculated to one-loop order
(and consistently separated from the renormalization terms) in a
series of papers \cite{BI-TP-88/32,BI-TP-90-43}.
For doing these one-loop calculations, the tree level
Lagrangian has first to be
completed by gauge-fixing and ghost terms in the
standard way. It was of particular convenience to perform the
calculations in the so-called Landau gauge\footnote{
The Landau
gauge is distinguished by the fact that ghost-couplings with
scalars ${\scriptstyle (\vec \pi, \vec \sigma)}$ vanish.
Hence, ghost loops
do not contribute to induced interactions (they are completely
absorbed in renormalization). As a consequence, the induced
interaction Lagrangians are fully gauge invariant and not only
BRS-invariant (as they would be in other gauges).}.
It has turned out
thereby, that these expressions can be represented in terms of fully
gauge-invariant Lagrangians as one would expect.
The resulting expressions -- together with the corresponding coupling
strengths -- are presented in ref.~\cite{BI-TP-88/32}
(for fermion interactions) and ref. \cite{BI-TP-90-43} (for bosonic
self interactions), respectively. Note that -- due to our
parametrization of the NL realization -- the scalar fields
$\pi$ and $\sigma$ emerge
always in the exponents. Consequently, the individual interaction
terms are obtained by expanding in powers of $\vec \pi$ and $\vec
\sigma$ and of the (unmixed) gauge bosons $\vec {\tilde{W^\mu}},
{\tilde Y}^\mu, \vec {\tilde{V^\mu}}$. These terms
are, of course, not individually symmetric -- only their sum is.\par
The main features of the resulting induced interactions can be
described as follows:
\begin{itemize}
\item{-} The strengths of all one-loop induced couplings are
logarithmically dependent of the cut-off $\Lambda$ (which we
sometimes have identified with a heavy-Higgs-mass $M_H$). Note that
higher loop contributions (though involving higher powers of $\Lambda$)
will not be dominant as long as 
$\Lambda \stackrel{<}{\sim} 3 \ {\rm TeV}$ \cite{YTP-80-01}, 
at least as long as one sticks to the interpretation of the nonlinear
theory as the limiting case of the linear one with $M_H \to \infty$.
\item{-} Quantum corrections to fermionic couplings of vector bosons
\cite{BI-TP-88/32} are -- although existing -- suppressed by a factor of 
$(m_f/M_W)^2$, and thus are negligible for light fermions (in
particular for electrons). Therefore, we can safely forget them for
our present purposes.
\item{-} There is no similar suppression for the vector boson self
interactions. In fact, the strengths of these additional interactions
are proportional to polynomials in $\lambda^2$ (of third power in
$\lambda^2$ for cubic self interactions, of fourth power in $\lambda^2$
for quartic ones) and therefore increase with increasing V-mass.
This is the aforementioned
manifestation of the non-decoupling nature of $M_V$
when $\lambda \to \infty$.
As to the specific structure of these interactions, an
interesting difference between cubic and quartic terms emerges
(cf. Tables 5 and 7 of ref.~\cite{BI-TP-90-43}): 
the cubic self-interactions have
the pure Yang-Mills structure\footnote{
Apart from the terms
proportional to
${\scriptstyle \partial_\mu G^\mu \scriptstyle (G = \vec
{\tilde W}, \tilde Y,
\vec
{\tilde V})}$.
They do not contribute to physical processes, since we have
to use consistently the Landau gauge, which yields transversal vector
propagators.}, whereas the quartic ones do not. This exceptional role
of the cubic interaction can be traced back to the fact that
nonrenormalizability is inferred to the BESS model via the
NL $\sigma$-model \cite{UT-KOMABA 76-12}.\par
In ref.~\cite{BI-TP-90-43} all vector boson self-interaction Lagrangians
are expressed in terms of unphysical (unmixed) vector fields
$\vec {\tilde W}, \tilde Y, \vec {\tilde V}$. For calculating
physical processes we need the corresponding expressions for the
physical fields $W^\pm, Z, A, V^\pm, V^0$, which can be obtained by
appropriately applying the mixing matrices. The resulting expressions
are quite lengthy. We summarize them in a fairly compact form (for
all interesting vertices) in App. A and B. Similarly, the induced
couplings of physical vector bosons to the (unphysical) scalar
fields $\vec \pi$ and $\vec \sigma$ have been calculated, since
they will be used in computing some amplitudes for
three-boson-production processes (see Fig. 1b), but we won't quote
them here explicitly.

Note that the approach of refs.~\cite{BI-TP-88/32} and \cite{BI-TP-90-43}
to separate systematically the cutoff-dependent one-loop effects into 
nonobservable (``renormalization'') and observable effects is based
on the interpretation of effective theories as first clearly promoted
by Appelquist {\it et al.\/} \cite{YTP-80-01}. However, other
interpretations and handling of effective theories may enjoy equal
legitimacy. For example, Casalbuoni {\it et al.\/}
\cite{UGVA-DPT-1986-01-492,CERN-TH-4876/87,hep-ph/9303201}
take the point of view that there are no quantum-induced terms
in the BESS effective theory, but that all terms should be treated
as phenomenological parameters. In addition, these authors consider
for the BESS the same one-loop radiative corrections as in SM and
equate $\Lambda = M_H \sim 1$ TeV. This interpretation leads to a somewhat
different BESS model than the BESS model discussed here, with the
differences ocurring at the one-loop level.


\section{Phenomenology}
Present $e^+e^-$-colliders like LEP I  allow only direct tests
of the couplings
between gauge bosons and fermions.
But, as we have shown, the nonrenormalizable
structure of the BESS model shows up most drastically
in the self-couplings of the gauge
bosons due to the new induced couplings.
Future $e^+e^+$-colliders with energies
above the $W^+W^-$ threshold (161 GeV)
will allow direct tests
of these self-couplings.
The first machine
to make the W-pair production process $e^+e^- \to W^+W^-$
possible
will be LEP II ($\sqrt s \sim 190 \ {\rm GeV}$),
but due to its very
limited energy range, deviations from the standard model will hardly be
observable.
A planned $e^+e^-$ (linear)
collider at $\sqrt s=500 \ {\rm GeV}$ (NLC)
will allow much more precise measurements of
the $e^+e^- \to W^+W^-$ cross section and, in particular,
a much better discrimination
of the BESS model
because of the expected higher integrated luminosity
of $20 fb^{-1}$ per year \cite{Leenen}
and because the CM energy of $500 \ {\rm GeV}$ is much larger
than the threshold of this reaction,
so that at this energy the violation of the
gauge cancellations due to the induced couplings in the BESS model
yields much higher deviations from the
standard model than at LEP II energies.
Furthermore,
 three gauge boson production processes like
$e^+e^- \to W^+W^-Z$ and $e^+e^- \to W^+W^-\gamma$ \cite{MAD/PH/420,UCD-88-24},
which supply a direct test of the quartic vector boson self-couplings
and the non-Yang--Mills structure of these
in the BESS model, will be measurable.
The $e^+e^- \to W^+W^-Z$ threshold is at $250 \ {\rm GeV}$ and the
high luminosity will enable even the
measurement of very small cross sections
in the order of $50fb$ as they are
expected for these processes.\par
Future hadron colliders and $\gamma\gamma$-collisions
at NLC will as well
supply tests of vector boson
self-interactions and of the induced couplings in
the BESS model, but these are not considered in the present analysis.
\par
In this paper we present the cross sections for the
two and three gauge boson
production processes at NLC energy of $\sqrt s=500 \ {\rm GeV}$.
In addition, we give an outlook
to what happens at an energy of $\sqrt s=2000 \ {\rm GeV}$,
which may be interesting for
machines of future generations.
\par
Specifically, we have calculated the following observables:
\item{-} Total cross sections for the
reactions
$e^+e^- \to W^+W^-, e^+e^- \to W^+W^-Z$ and $e^+e^-\to W^+W^-\gamma$
both for polarized and
nonpolarized gauge bosons.
\item{-} The following partial cross sections (distributions):
\begin{itemize}
\item{$\bullet$} $d\sigma/d \cos \theta_G$ for two and
three gauge boson production $(G = W, Z, \gamma)$,
\item{$\bullet$} $d\sigma/dE_G$ for three
gauge boson production,
\item{$\bullet$} $d\sigma/dP_{G,T}$
for three gauge boson production,
\item{$\bullet$} $d\sigma/dy_G$  for three gauge boson production.
\end{itemize}
\item{-}For the reaction $e^+e^- \to ZZZ$
we have calculated only the total cross section,
since distributions will
presumably not be measurable because of the small size of the cross
sections.\par
\end{itemize}
To identify the gauge boson production events in experiment, one has
to reconstruct the W and Z bosons from their decay products, while
the photons can be identified directly. As Frank, M\"attig,
Settles and Zeuner \cite{FMSZ} have stated, 
the most significant $e^+e^- \to W^+W^-$
events are those, where one W decays into leptons and the other
one into hadrons. An analysis of the angular distribution of the
decay products of the W and Z bosons yields the polarisation of these
bosons, so that the cross section for the production of polarized
gauge bosons can be measured, too.\par
Figures 1a and b show schematically the tree level Feynman diagrams
which contribute to the two and three gauge boson production processes.
(Since we performed our calculations in the Landau gauge, there is
one diagram with an exchange of an unphysical would-be Goldstone
boson.)
We formally calculated the cross sections at tree level, but for
each gauge
boson self-interaction vertex we took into account
the one-loop induced coupling.
All cross sections acquire
contributions from cubic self-couplings, whereas in the three
gauge boson production processes even quartic self-couplings
get involved due
to the last diagram. In the case of $e^+e^- \to ZZZ$ there
is a diagram with a coupling
of four neutral gauge bosons which does not exist for
pure Yang-Mills type
interactions, but exists in the BESS model as a consequence
of the violation of the Yang-Mills structure.
\par
To calculate the cross section for the W pair production we have
proceded completely analytically
using the usual trace techniques. For the calculation of the three
gauge boson production cross sections we had to
procede numerically.
We calculated the amplitudes using
helicity techniques \cite{MAD/PH/420,DESY 85/133} and integrated
numerically over the phase space.
In agreement with \cite{MAD/PH/420} 
we imposed the following transverse-momentum and
pseudorapidity cuts on the photon produced in
$e^+e^- \to W^+W^-\gamma$:
$$
P_{T,\gamma}>20 \ {\rm GeV} ,~~~~\vert \eta_\gamma \vert < 2 .\eqno(3.1)
$$
In order to obtain numerical values for the
cross sections in the BESS model, we had
to specify the free parameters of the model, i.e.,
$g$,$g'$ and $g''$\footnote{
Bear in mind, that ${\scriptstyle g}$ and ${\scriptstyle g'}$
are neither
physically nor numerically identical
with the standard model ${\scriptstyle g}$ and ${\scriptstyle g'}$,
since ${\scriptstyle g\cos\varphi}$ and not ${\scriptstyle g}$ describes
the coupling of the ${\scriptstyle W}$ bosons to fermions.},
$f^2$, $\lambda^2$ and the cut-off $\Lambda$. $\Lambda$ was set to
$$
\Lambda = 5 \ {\rm TeV} \eqno(3.2)
$$
We are aware that this value may be too large, because the two-loop
contributions may become dominant for such a choice, as mentioned before.
However, a different choice of $\Lambda$, e.g., $\Lambda=2$ TeV, would change
the values of the deviations from SM by only a few per cent. This 
is due to the fact that only parts of these deviations are due to
the one-loop induced couplings, while the larger part stems from tree level 
effects (existence of heavy gauge bosons, gauge boson mixing).
The other five parameters can be determined from
chosen values of $\alpha_{\rm em}$, $M_W$, $M_Z$, $M_{V_0}$ and $g''/g$.
The first three values are empirically given (the electromagnetic
fine structure
constant $\alpha_{\rm em}$ was taken as 1/127,
which is the value of the running coupling
constant at the $e^+e^- \to W^+W^-Z$
threshold of 250 GeV\footnote{
The slight
numerical differences between our predictions and the results of
\cite{MAD/PH/420,UCD-88-24}
might be traced back to a slightly
different choice of the coupling parameters.
Different choices for ${\scriptstyle \alpha_{\rm em}}$ would result
approximately only in multiplying in SM and BESS cross sections
by a common factor close to 1, an effect
which would not appreciably change
the deviations of BESS cross sections from those of SM.\vfill}.)
while the free parameter $g''/g$ was set to
$$
g''/g = 10,\eqno(3.3)
$$
a value suggested by the apparent success of SM in fermionic
processes.
For the unknown mass $M_{V^0}$ we chose the following reference values:
$$
M_{V^0}=400 \ {\rm GeV},~~1000 \ {\rm GeV},~~2000 \ {\rm GeV},
{}~~2500 \ {\rm GeV},\eqno(3.4)
$$
which indicate the presumable range within which $M_{V^0}$
is expected to lie.
Remember that $\lambda^2$ grows with $M_{V^0}$, so a heavy $\rm V^0$
means
strong induced couplings.
\par
As a further input for our calculation
we need the $\rm V^\pm$ and $\rm V^0$
widths. We calculated these widths taking the induced couplings
fully into account. The main
decay channels are two and three gauge bosons and fermion pairs.
This will be discussed in detail in a forthcoming paper \cite{BI-TP-90/14}.
Here in Table \ref{t1} we only present
the results for our reference values.
Note that for both $\rm V^\pm$ and $V^0$, the width is dominated by the
two-vector channels and increases strongly with $M_V$,
such that masses of
$M_V \stackrel{>}{\sim} 2200 \ {\rm GeV}$ are unlikely.
In this respect
the reference mass
$M_{V^0}= 2500 \ {\rm GeV}$ has to be considered as an extreme case.
\par
\begin{table}[htb]
\caption{Widths of the $V$ bosons.}
\label{t1}
\begin{ruledtabular}
\begin{tabular}{lllll}
$\lambda^2$ & $M_{V^0} ({\rm GeV})$ & $\Gamma_{V^0} ({\rm GeV})$ & $M_{V^\pm} ({\rm GeV})$ & $\Gamma_{V^\pm} ({\rm GeV})$ \\  \hline
0.241 & 400 & 0.829 & 399.5 & 0.748 \\
1.505 & 1000 & 31.17 & 999.0 & 38.54 \\
6.012 & 2000 & 2300 & 1997.3 & 1138 \\
9.406 & 2500 & 31197 &     2496.7 & 23093
\end{tabular}
\end{ruledtabular}
\end{table}
For comparison we calculated the cross sections for the standard
model as well. Thereby
we neglected all Higgs boson effects, which makes sense if the Higgs
boson is lighter than $2M_W$ or heavier than
$\sqrt s - M_Z$ because then the Higgs boson
yields only a negligible contribution. Else, the
Higgs boson would show up as a
resonance in the $W^+W^-$
channel and can be identified by a Jacobian Peak
 in the energy spectrum of the produced $Z$ bosons.
So Higgs boson effects can
easily be distinguished from
effects of the induced couplings in the BESS model.

One further remark is in order here: In calculating the various 
cross sections in the way described above (tree level diagrams 
including one-loop induced couplings) we have not included the 
finite (cutoff-independent) one-loop corrections, although they
are not much smaller in size than the induced (cutoff-sensitive)
contributions, at least as long as the cutoff is of reasonable size 
(1-5 TeV). However, these cutoff-independent corrections are expected to
be almost the same as the corresponding finite (one-loop) corrections
within SM (calculated within that specific renormalization scheme
which corresponds to the replacement $1/(4 - d) \mapsto \ln (\Lambda/M_W)
= \ln(M_H/M_W)$). The only difference stems from the diagrams with heavy boson
lines and is numerically suppressed relative to the typical SM
corrections by (powers of) $(M_W/M_V)^2$ or $(M_W/M_H)^2$.
Therefore, the difference between the BESS and the SM results 
(at one-loop level) is practically insensitive to the finite corrections
since they are almost the same for both.

We note here that the production of the BESS resonances at $e^+ e^-$
linear colliders has recently been studied also by Casalbuoni {\it et al.\/}
\cite{hep-ph/9303201}. Furthermore, an analysis in the latter work,
based on the LEP data communicated at the Dallas Conference (1992),
yields the free parameter $(g^{\prime\prime}/g) \geq 14-17$, depending on the mass
of the top quark being 180 or 150 GeV (when their phenomenological parameter
$b$ is set equal to zero). Choosing these larger values for $(g^{\prime\prime}/g)$,
instead of the chosen $(g^{\prime\prime}/g) = 10$, the deviations from SM
calculated in the present paper would certainly become smaller. However,
we have to stress that the BESS model considered and analyzed by the
authors of ref.~\cite{hep-ph/9303201} is somewhat different from the
BESS model considered here; above all due to the different treatment and
interpretation of the quantum one-loop effects, 
as already mentioned at the end of Sec.~II. Actually, 
the authors of  ref.~\cite{hep-ph/9303201} treat the coefficients of the
possible (induced) terms as unspecified parameters, and in their
detailed analysis they set most of them equal to zero. We believe that,
when incorporating {\it all\/} possible parameters as freely adjustable,
the resulting bounds on  $(g^{\prime\prime}/g)$ would be considerably different.
In any case, it is our experience that the bounds on $(g^{\prime\prime}/g)$,
when determined within our scenario (i.e., by taking our values for the
indicated parameters), turn out to be different 
(see ref.~\cite{BI-TP-90-40}). Therefore, we do not take the values of 
ref.~\cite{hep-ph/9303201} as stringent.

\section{Discussion of Numerical Results}
The results for the different observables (cross sections,
distributions, asymmetries) are plotted in Figures 2-21. In all
figures the solid lines represent the results for the Higgsless
standard model and the different types of broken lines the
predictions of the BESS model for different choices of $M_{V^0}$, i.e.,
for different strengths of the induced couplings.

\subsection{${\bf W^+W^-}$-Production}
In Figs. 2-5 we have depicted the resulting values of the total
cross section for $e^+ e^- \to W^+W^-$ respectively by the BESS
model
(together with the SM-predictions)
for different polarization states of the final vector bosons
(unpolarized, transversally (TT), longitudinally (LL) and mixed
polarized (LT + TL) $W$'s). We plot the cross sections for two
energy intervals:
$$
{\rm (a)} \;\:  \sqrt s = 150-550 \ {\rm GeV} \eqno(4.1)
$$
(which covers the energy region to be reached by the planned NLC)
and
$$
{\rm (b)} \;\; \sqrt s = 0 - 2500 \ {\rm GeV} \eqno(4.2)
$$
(which roughly represents the total energy region where the BESS
model is reasonably assumed to work).\par
We have used the $M_{V^0}$-values as specified in ch. 3 ($M_{V^0} = 400,
1000, 2000, 2500 \ {\rm GeV}$) for the wide energy range (4.2), whereas,
for clarity of presentation, only three $M_{V^0}$-values
($M_{V^0} = 400, 1000, 2500 \ {\rm GeV}$) were used for the narrow (NLC-)
energy range a).\par
Differential cross sections $d \sigma/d \cos \theta$
at $\sqrt s = 500 \ {\rm GeV}$, for non-polarized, (LT- and TL-polarized
and LL-polarized $W$'s are depicted in Figs. 6-8 for the angular
range $-1 \le \cos \theta \le +1$ (a). In each case, we isolated
in addition the forward direction ($0.9 \le \cos \theta \le 1$)
(b) since, in general, the cross sections are particularly large in
this region and, furthermore, the deviations from SM show a
substantial variation there. For comparison, we also quoted
$d\sigma /d \cos \theta$ at $\sqrt s = 2000 \ {\rm GeV}$ (Fig. 9) for
unpolarized vector bosons and for the same two angular regions.
The related forward-backward asymmetries $A_{FB}$ and centre-edge
asymmetries $A_{CE}$ for unpolarized $W$'s are depicted in Figs. 10
and 11, again for the same two energy ranges as for the total
cross sections.\par
Let us now comment on all these results, in particular on the
differences between SM and BESS model predictions.\par
The deviations from SM stem from three effects:
\begin{itemize}
\item{a)} exchange of heavy boson $V^0$ (including the resulting
resonance effects),
\item{b)} mixing between light and heavy bosons,
\item{c)} deviations of the $Z^0 W^+ W^-$ and $\gamma W^+W^-$
couplings strengths from those of SM due to induced interactions
which result in violation of gauge cancellation between $t$ and
$s$ channels.\par
\end{itemize}

For $\sqrt{s} = 500 \ {\rm GeV}$ and $M_V$ in the range
$700 \ {\rm GeV} \stackrel{<}{\sim} M_V \stackrel{<}{\sim} 1500 \ {\rm GeV}$, the
largest contribution comes from the tree level effects (a) and (b).

For $M_{V^0} \stackrel{<}{\sim} 1000 \ {\rm GeV}$, the $V^0$-resonance is narrow and
pronounced. For heavy $V^0$ ($M_{V^0} \stackrel{>}{\sim} 2000 \ {\rm GeV})$, the
$V^0$-resonance peak becomes very broad, in fact invisible, but
the induced couplings are then large and yield large deviations from
SM. For instance, at $\sqrt s = 500 \ {\rm GeV}$, the relative deviations of
the total cross section (for non-polarized $W$'s) are 3 \%, 5\%, 8\%,
17 \% for $M_{V^0} = 400, 1000, 2000, 2500 \ {\rm GeV}$, respectively. Such
deviations should therefore be observable with a $500~{\rm GeV}~e^+e^-$
collider reaching a luminosity of $20~fb^{-1}$ per year\footnote{
If one assumes only a luminosity of $10 \ {\rm fb}^{-1}$ per year
\cite{FMSZ}, all the statistical errors get enlarged by a factor
of $\sqrt{2}$ in the following discussion.} 
where an experimental error smaller than 3 \% should be possible (the
statistical error for 90 \% confidence level being very small
$(\sim 0.8 \%)$ for a one year's collection of data.\footnote{
The reduction factor 0.3
was taken into account for the really
reconstructable events, as suggested by Frank, M\"attig, Settles 
and Zeuner\cite{FMSZ}.} 
These relative deviations are much more pronounced
for L-T polarized $W$'s (5,7 \%, 11.3 \%, 34 \%, 104 \%, respectively;
statistical error is about 3.5 \%) and especially for L-L
polarized $W$'s\footnote{
For the production of polarized vector bosons, the absolute
statistical error is to a good approximation equal to the one
corresponding to the production of unpolarized bosons \cite{Prep2},
which yields a larger relative error.} 
(-43 \%, 44 \%, 164 \%, 560 \%,
respectively; statistical
error is about 5.5 \%). On the other hand, for T-T polarized  $W$'s, the
deviations are practically independent of $M_{V^0}$ (about 4 \%;
statistical error being roughly equal to that of the case of
nonpolarized W's).\par
{}From Figs. 10 and 11 we see that the deviations of $A_{FB}$ and
$A_{CE}$ at $\sqrt s = 500 \ {\rm GeV}$ are small ($\sim 1 - 2 \%$ for
$M_{V^0} < 1000 \ {\rm GeV}$) but will increase drastically with higher
energies.\par
Figs. 6a,b show that the relative deviations of $d \sigma/d \cos
\theta$ from SM values at $\sqrt s = 500 \ {\rm GeV}$
for non-polarized $W$'s are substantial for negative values of
$\cos \theta$ (e.g. at $\cos \theta = - 0.5$ they are - 14 \%, + 18 \%,
+ 200 \% for $M_{V^0} = 400, 1000, 2500 \ {\rm GeV}$). However, the absolute
values of the cross sections are very small at such angles and the
statistical errors are therefore large ($\sim 14.5 \%$ for
$\Delta (\cos \theta) \approx 0.1$). On the other hand, the
deviations in the forward region $(\cos \theta = 0.9 - 0.99)$ are
approximately 4 \% - 6 \% for any $M_{V^0}$. The statistical error
here is small (about 1 \% for $\Delta (\cos \theta) = 0.09$). Hence,
it appears that it may be more promising to measure $\Delta
(d \sigma/d \cos \theta)_{NP}$ in the
forward directions than in other directions. The relative
deviations are substantially larger for LT + TL channel ($\sim
30 - 45 \%$ for $\cos \theta = -0.5$ and $\sim 1.5 \%$ for $\cos
\theta = 0.9,~M_{V^0} = 400, 1000 \ {\rm GeV}$). However, the corresponding
statistical errors under the mentioned conditions are also large
($\sim 45 \%$ and $10 \%$, respectively) due to small absolute values.
These features (large deviations from SM but large small total
cross-sections and large statistical errors) are even more pronounced
in the case of LL polarization. Hence it appears that differential
cross sections for polarized W's are not particularly useful
quantities for discriminating modes, due to large statistical errors.
\par
As seen from Figs. 9 a and b, $(d \sigma/ d (\cos \theta))$
values are drastically  decreased at high energies $(\sqrt s =
2000 \ {\rm GeV})$.
\subsection{Three Gauge Boson Production}
Figs. 12-14 show the total cross sections for the reactions
$e^+e^- \to W^+W^-Z,~e^+e^- \to W^+W^-\gamma$ and $e^+e^- \to ZZZ$
(all vector bosons unpolarized) as functions of the total energy
$\sqrt s$, where $\sqrt s$ varies again in the two ranges (4.1)
and (4.2). Let us first discuss the larger region (4.2) which shows
the global behaviour of the cross sections. In addition to the
resonance peaks at $\sqrt s = M_{V^0}$ (due to the exchange of a
$V^0$ boson in the $s$ channel\footnote{
In the reaction ${\scriptstyle e^+e^-
\to ZZZ}$ there are no visible resonances, since if the
${\scriptstyle V}$ bosons are
light the induced couplings of four neutral gauge bosons which are
responsible for these resonances are too small and if the
${\scriptstyle V}$ bosons
are heavy the resonances are too wide.}) which also occur in two body
productions, there are $V^0$ resonances in the $W^+W^-$ sub-channel
and $V^\pm$ resonances in the $W^\pm Z$ sub-channel (see Fig. 16)
(for $e^+e^- \to W^+W^-Z$ and similar for $e^+e^- \to W^+W^- \gamma$,
but not for $e^+e^- \to ZZZ$, because there are no diagrams with
trilinear couplings). This means that e.g. the direct $e^+e^- \to
W^+W^-Z$ reaction becomes superimposed by the reactions $e^+e^- \to
V^0Z$ with consequent decay $V^0 \to W^+W^-$ and by the reaction
$e^+e^- \to V^\pm W^\mp$ with the decay $V^\pm \to W^\pm Z$. So above
the threshold of these reactions, i.e., at slightly larger energies
than the $V^0$ resonance, the cross section shows again a maximum. When
$M_{V^0}$ gets larger, all $V$-resonances broaden (see table 1) and
finally dissapear when the $V$-boson width is in the order of or
greater than the $V$ boson mass.\par
An important effect of the induced couplings, as in the case of
$W$ pair production, consists in destroying the gauge cancellations,
i.e., the parts of the amplitudes for the different graphs which grow
with energy do not cancel completely anymore as they would do in
a renormalizable gauge theory. This leads to deviations of the BESS
model cross sections from the standard model ones, which grow both
with energy and with $M_{V^0}$, (since $M_{V^0}$
is proportional to the
strength of the induced couplings). Note that this effect is more
pronounced for triple boson production as compared to $W^+W^-$
production since the induced quartic couplings go with a higher power
of $\lambda^2$.\par
At NLC energies (4.1) the deviations from SM are not so drastic,
because the non cancelled parts of the amplitudes, which grow with the
energy, are still not so big. Except for the case of a very light
$V$-boson, which causes resonances at low energies, there are only
deviations in the per cent region. (In case of $e^+e^- \to W^+W^-Z$
from 8 \% for medium $M_{V^0}$ up to 20 \% for heavy $M_{V^0}$.)
However, these deviations are large enough to be measurable. The
$e^+e^- \to W^+W^-Z$ cross section is about 50 fb. With an expected
NLC-luminosity of $20 fb^{-1} a^{-1}$ there are 1000 annual events.
Following the analysis of Barger, Han and Phillips 
\cite{MAD/PH/420}, 20 \% of
them, that means 200 events per year, will be reconstructable, which
means that the statistical error (for 90 \% confidence level) can
be suppressed to $\sim 5 \%$ after five years of run. The systematical
error is expected to be 2 \% . Thus it should in principle be
possible to distinguish the BESS model with the given parameters from
the standard model empirically\footnote{
It should be mentioned
that most of the deviations from the standard model for medium
${\scriptstyle M_{V^0}}$ at NLC
energies are tree level effects and not caused by
induced couplings. So an empirical verification of the BESS model
does not neccesarily imply a verification of the induced couplings.}.
The same is true for the reaction $e^+e^- \to W^+W^-\gamma$. On the
other hand, the cross section for the reaction $e^+e^- \to ZZZ$ is
only 1fb, which means the statistical error is probably too large
to get precise results, in reasonable time, although this reaction
would be of largest importance because of its singular nature.\par
Figure 15 shows the cross sections for the production of polarized
gauge bosons in the reaction $e^+e^- \to W^+W^-Z$. The differences
of the BESS model to the standard model are small if no or only one
longitudinally polarized gauge boson is produced and they are large
if two or three longitudinally polarized gauge bosons are produced.
This is because the amplitudes of the single Feynman graphs grow with
higher powers of $\sqrt s$ the more longitudinal bosons are in the
final state, so the effect of non-cancellation of the leading powers
of $\sqrt s$ is especially strong if mainly longitudinal gauge bosons
are produced. Unfortunately, in those cases the total cross sections
are very small and so the statistics are very bad, while if
transversal gauge bosons are produced, the statistics are better
because of the larger cross sections but the deviations are small.\par
Figures 16 to 21 show different partial cross sections at the NLC
energy of 500 GeV. Except for resonance effects (in case of a light
$V^0$) like Jacobian peaks\footnote{
There are only very small
effects of a ${\scriptstyle V^\pm \to W^\pm \gamma}$
resonance in the ${\scriptstyle e^+e^-
\to W^+W^- \gamma}$ process since the responsible coupling vanishes
on tree level and the induced coupling is very small.} the
deviation of the BESS model from the SM results are distributed
regularly over the angular, energy etc. region. Thus, a measurement
of only those processes where one of the produced gauge bosons is
emitted in a certain part of its phase space would not improve
the expected deviation from the standard model, but it would make
statistics worse because there are less events.

\section{Conclusions}
The discussion of the last section has shown that the specific
structures of the BESS model (existence of heavier vector bosons,
mixing between heavy and light ones, new induced couplings) will
become effective in boson production by $e^+e^-$-collisions at
energies of about 500 GeV with sufficient magnitude, such that an
identification of these effects (and, consequently discrimination
of BESS and SM) would be possible with the help of the planned New
Linear Collider (NLC). The most promising observables in this respect
are total cross sections for production of unpolarized and
longitudinally polarized gauge bosons. Further valuable information
can also be obtained by measuring asymmetries and partial cross
sections although the results will be less conclusive due to limited
statistics.\par
In general, quantities connected with two boson production will be
measurable to much greater accuracy and yield more distinctive
bounds. But results for three boson production processes will be
of particular theoretical interest because they are determined
partially by the four-boson self-interactions which are much more
sensitive to the specific model than the three-vector vertices.\par
If no deviations from SM will be found in future measurements
of the above-mentioned process the results will nevertheless allow
to restrict the parameter ranges of the BESS model parameters due to
finite experimental accuracy. A careful investigation of the
corresponding expectations has been published in ref.~\cite{hep-ph/9212291}.

\begin{acknowledgments}
\noindent
This work was supported in part by the Deutsche Forschungsgemeinschaft (DFG),
Project No.~ Ko 1062/1-2.
\end{acknowledgments}

\appendix

\section{Trilinear gauge boson self-interactions}
\label{appA}

The one-loop induced interaction Lagrangians containing three
unphysical (unmixed) vector boson fields\footnote{
see footnote 4} are written down in Table 5 of ref.~\cite{BI-TP-90-43}. Here we
can consistently forget about the terms proportional to $\partial_\mu
G^\mu~~~(G = \vec W, Y, \vec V)$ since we work in the Landau gauge.
Decomposing the remaining expressions into charge-eigenstates,
applying the mixing matrices ${\cal C}$ (for charged vector bosons)
and ${\cal N}$ (for neutral ones) (cf. 2.2 and 2.3) and adding the
corresponding tree level interactions we get the total Lagrangian
for the cubic self interactions of physical vector bosons (tree level
and one-loop induced ones). It can be written in a compact form by
using the notation
$$
{W^\pm \choose V^\pm} \equiv  {c_1^\pm
\choose c_2^\pm} \ , ~~~~~~~~~~~~~~
\left( 
\begin{matrix}
A \cr
Z \cr
V^0\cr
\end{matrix} 
\right) \equiv 
\left( 
\begin{matrix}
n_1 \cr
n_2 \cr
n_3 \cr
\end{matrix}
\right) \ . \eqno(A.1a,b)
$$
One obtains
\bea*
{\cal L}_{\rm 3 gauge~bosons} &=& i \sum^2_{a,b =1} \sum^2_{i=0}
K_{a b, i} \cdot \{ (c_a^+)^\mu (c_b^+)^\nu (n_i)_{\mu\nu} +
+ [(c_a^-)_{\mu \nu} (c_b^+)^\mu - (c_b^+)_{\mu\nu} (c_a^-)^\mu]
(n_i)^\nu\}
\qquad\qquad\qquad\qquad    (A.2)
\eea*
where
\bea*
K_{a b, i} &=& 2 {\cal C}_{1a} {\cal C}_{1b} [\alpha_1 {\cal N}_{oi}
+ \alpha_6 {\cal N}_{1i} + \alpha_2 {\cal N}_{2i}] + 
 2 {\cal C}_{2a} {\cal C}_{2b} [\alpha_3 {\cal N}_{oi} + \alpha_4
{\cal N}_{1i} + \alpha_7 {\cal N}_{2i}] + \cr
\nonumber\\
&&+ ({\cal C}_{1a} {\cal C}_{2b} + {\cal C}_{2a} {\cal C}_{1b})
[2 \alpha_2 {\cal N}_{1i} + 2 \alpha_4 {\cal N}_{2i} + \alpha_5
{\cal N}_{oi}]
\qquad \qquad \qquad \qquad \qquad \qquad \qquad \qquad \;\;\; (A.3)
\eea*
and [we use the notation: $t \equiv [48 (4 \pi)^2]^{-1} \ \ln(\Lambda/M_W)$]
\bea*
\alpha_1 &=& - t {1 \over 4} (\lambda^2 - 1) (\lambda^2 - 3)^2\cr
\alpha_2 &=& t {1\over 2} (\lambda^6 - 3 \lambda^4 + 5 \lambda^2 + 1)\cr
\alpha_3 &=& - t (\lambda^6 - 2 \lambda^4 - 1)\cr
\alpha_4 &=& \alpha_3\cr
\alpha_5 &=& t (\lambda^6 - 5 \lambda^4 + 3 \lambda^2 + 1)\cr
\alpha_6 &=& - t {1 \over 4} (\lambda^6 + \lambda^4 - 5 \lambda^2 + 11)
+ {1 \over {2 g^2}}\cr
\alpha_7 &=& t \cdot 2 (\lambda^6 - 2 \lambda^4 - 1) + {2 \over
{g^{\prime\prime 2}}}
\qquad \qquad \qquad \qquad \qquad \qquad \qquad \qquad \qquad (A.4)
\eea*

\section{Quadrilinear gauge boson self-interactions}
\label{appB}

The one-loop induced interaction Lagrangians connecting four (unmixed)
vector boson fields can be found in Table 7 of ref.~\cite{BI-TP-90-43}. 
They can be
inverted into expressions for quartic interactions of the physical
bosons by appropriately using the mixing matrices ${\cal C}$ and
${\cal N}$ (cf. (2.2) and (2.3)). Here, we refrain from quoting the
full Lagrangian but we list only quartic (physical) vector boson
self-couplings (tree level + one-loop induced ones) which contribute to
the processes $e^+e^- \to W^+W^-Z,~~e^+e^- \to W^+W^-\gamma$ and
$e^+e^- \to ZZZ$.
\bea*
{\cal L}_{W^\pm W^\pm \gamma\gamma} =& \sigma_1 (A^\mu A_\mu)
(W^{+\nu}W^-_\nu) + 2 \sigma_2 (A^\mu A^\nu) (W^+_\mu W^-_\nu) 
&(B.1a)\cr
\nonumber\\
{\cal L}_{W^\pm W^\pm \gamma Z} =& \sigma_3 (A^\mu Z_\mu)
(W^{+\nu}W^-_\nu) 
+ \sigma_4 (A^\mu Z^\nu) (W^+_\mu W^-_\nu + W^+_\nu W^-_\mu) 
&(B.1b)\cr
\nonumber\\
{\cal L}_{W^\pm W^\pm \gamma V^0} =& \sigma_5 (A^\mu V^0_\mu)
(W^{+\nu}W^-_\nu) + \sigma_6 (A^\mu V^{0 \nu})
(W^+_\mu W^-_\nu + W^+_\nu W^-_\mu) 
&(B.1c)\cr
\nonumber\\
{\cal L}_{W^\pm W^\pm ZZ} =& \sigma_7 (Z^\mu Z_\mu)
(W^{+\nu}W^-_\nu) + 2 \sigma_8 (Z^\mu Z_\mu) (W^+_\mu W^-_\nu) 
&(B.1d)\cr
\nonumber\\
{\cal L}_{W^\pm W^\pm ZV^0} =& \sigma_9 (Z^\mu V^0_\mu)
(W^{+\nu}W^-_\nu) + \sigma_{10} (Z^\mu V^{0 \nu})
(W^+_\mu W^-_\nu + W^+_\nu W^-_\mu)
&(B.1e)\cr
\nonumber\\
{\cal L}_{ZZZZ} =& \sigma_{11} (Z^\mu Z_\mu) (Z^\nu Z_\nu)
&(B.1f)\cr
\nonumber\\
{\cal L}_{ZZZV^0} =& \sigma_{12} (V^{0\mu} Z_\mu) (Z^\nu Z_\nu)
&(B.1g)\cr
\eea*
It turns out that all couplings of four neutral gauge bosons where
at least one of these is a photon are zero. Note also that there are
(non-vanishing) interaction terms involving photon fields which
individually are not invariant under $U(1)_{em}$. However, since
they are obtained from fully invariant expressions (by expansion in
power of fields), electromagnetic gauge invariance is established
if the appropriate terms (including in general also cubic boson
interaction terms) are added.\par
The coupling constants are given by:
\bea*
\sigma_1 &= & 2
\{ \lbrack
\beta_2 {\cal N}_{00} {\cal N}_{10} +
\beta_4 {\cal N}_{10} {\cal N}_{20} + 
\beta_{11} {\cal N}^2_{00} + \beta_{15} {\cal N}_{20}^2 
+ \beta_{24} {\cal N}_{00} {\cal N}_{20} + 2 \beta_{40}
{\cal N}_{10}^2
\rbrack
{\cal C}^2_{11} 
\nonumber\\
&& +
\lbrack
\beta_4 {\cal N}^2_{10} + \beta_8 {\cal N}^2_{20} 
+ 2 \beta_{18} {\cal N}_{10} {\cal N}_{20} + \beta_{26} {\cal N}_{00}
{\cal N}_{10} + \beta_{29} {\cal N}_{00}^2 + \beta_{35} {\cal N}_{00}
{\cal N}_{20}
\rbrack
{\cal C}_{11} {\cal C}_{21} 
\nonumber\\
&&+
\lbrack
\beta_8 {\cal N}_{10} {\cal N}_{20} + \beta_{10} {\cal N}_{00}
{\cal N}_{20} + \beta_{15} {\cal N}^2_{10} + \beta_{20} {\cal N}^2_{00}
+ \beta_{33} {\cal N}_{00} {\cal N}_{10} + 2 \beta_{42}
{\cal N}^2_{20}
\rbrack
{\cal C}^2_{21}
\}
\qquad \qquad \;\;\;\;\;\;\;\;\;\;\;\;\;\;\;\;\;\;
\;\;\;\;\;\;\;\; (B.2a)\cr
\nonumber\\
\sigma_2 &= &
\lbrack
\beta_1 {\cal N}_{00} {\cal N}_{10} +
\beta_3 {\cal N}_{10} {\cal N}_{20} 
+ \beta_{12} {\cal N}^2_{00} + \beta_{16} {\cal N}_{20}^2 
+ \beta_{25} {\cal N}_{00} {\cal N}_{20} + 2 \beta_{39}
{\cal N}_{10}^2
\rbrack
{\cal C}^2_{11} 
\nonumber\\
&&+
\lbrack
\beta_3 {\cal N}^2_{10} + \beta_7 {\cal N}^2_{20} 
+ 2 (\beta_{17} + \beta_{19}) {\cal N}_{10}
{\cal N}_{20} + (\beta_{27} + \beta_{28})  {\cal N}_{00}
{\cal N}_{10} 
+ \beta_{30} {\cal N}_{00}^2 + (\beta_{36} + \beta_{37})
{\cal N}_{00} {\cal N}_{20}
\rbrack
{\cal C}_{11} {\cal C}_{21}
\nonumber\\
&&+
\lbrack
\beta_7 {\cal N}_{10} {\cal N}_{20} + \beta_9 {\cal N}_{00}
{\cal N}_{20} 
+ \beta_{16} {\cal N}^2_{10} + \beta_{21} {\cal N}^2_{00} 
+ \beta_{34} {\cal N}_{00} {\cal N}_{10} + 2 \beta_{41}
{\cal N}^2_{20}
\rbrack
{\cal C}^2_{21}  \qquad \qquad \; \;  
\;\;\;\;\;\;\;\;\;\;\;\;\;\;\;\;\;\;
\;\;\;\;\;\;\;\;\ (B.2b)\cr
\nonumber\\
\sigma_3 &= & 2
\{ \lbrack
\beta_2 ({\cal N}_{01} {\cal N}_{10} + {\cal N}_{00} {\cal N}_{11}) +
\beta_4 ({\cal N}_{11} {\cal N}_{20} + 
{\cal N}_{10} {\cal N}_{21})
+ 2 \beta_{11} {\cal N}_{00} {\cal N}_{01} +
2 \beta_{15} {\cal N}_{20} {\cal N}_{21}
+ 
\beta_{24} ({\cal N}_{01} {\cal N}_{20} + {\cal N}_{00}
{\cal N}_{21}) 
\nonumber\\ &&
+ 4 \beta_{40}
{\cal N}_{10} {\cal N}_{11}
\rbrack
{\cal C}^2_{11}  + \lbrack
2 \beta_4 {\cal N}_{10}{\cal N}_{11} + 2 \beta_8
{\cal N}_{20}{\cal N}_{21} 
+ 2 \beta_{18} ({\cal N}_{11}
{\cal N}_{20} + {\cal N}_{10} {\cal N}_{21}) +
\beta_{26} ({\cal N}_{01} {\cal N}_{10} + {\cal N}_{00}
{\cal N}_{11}) 
\nonumber\\
&&+ 
\beta_{29} {\cal N}_{00} {\cal N}_{01} + \beta_{35} ({\cal N}_{01}
{\cal N}_{20} + {\cal N}_{00} {\cal N}_{21})
\rbrack
{\cal C}_{11} {\cal C}_{21} 
+
\lbrack
\beta_8 ({\cal N}_{11} {\cal N}_{20} + {\cal N}_{10}
{\cal N}_{21})
\beta_{10} ({\cal N}_{01} {\cal N}_{20} + {\cal N}_{00}
{\cal N}_{21}) 
\nonumber\\ &&
+ 2 \beta_{15} {\cal N}_{10} {\cal N}_{11} +
2 \beta_{20} {\cal N}_{00} {\cal N}_{01} 
+ \beta_{33} ({\cal N}_{01} {\cal N}_{10} + {\cal N}_{00}
{\cal N}_{11}) + 4 \beta_{42}
{\cal N}_{20} {\cal N}_{21}
\rbrack
{\cal C}^2_{21}
\}
 \qquad \qquad \;\; \;\;\;\;\;\;\;\;\;\;\;\;\;\;\;\;\;\;
\;\;\;\;\;\;\;\; (B.2c) \cr
\nonumber\\
\sigma_4 &= &
\lbrack
\beta_1 ({\cal N}_{01} {\cal N}_{10} +
{\cal N}_{00} {\cal N}_{11}) +
\beta_3 ({\cal N}_{11} {\cal N}_{20} + {\cal N}_{10} {\cal N}_{21})
+ 2 \beta_{12} {\cal N}_{00} {\cal N}_{01} +
2 \beta_{16} {\cal N}_{20} {\cal N}_{21} 
+ \beta_{25} ({\cal N}_{01} {\cal N}_{20} + {\cal N}_{00}
{\cal N}_{21}) 
\nonumber\\ &&
+ 4 \beta_{39}
{\cal N}_{10} {\cal N}_{11}
\rbrack
{\cal C}^2_{11} 
+ \lbrack
2 \beta_3 {\cal N}_{10}{\cal N}_{11} + 2 \beta_7
{\cal N}_{20}{\cal N}_{21} 
+ 2 (\beta_{17} + \beta_{19}) ({\cal N}_{11}
{\cal N}_{20} + {\cal N}_{10} {\cal N}_{21}) 
\nonumber\\ &&
+ (\beta_{27} + \beta_{28}) ({\cal N}_{01}
{\cal N}_{10} + {\cal N}_{00}
{\cal N}_{11}) 
+ 2 \beta_{30} {\cal N}_{00} {\cal N}_{01} +
(\beta_{36} + \beta_{37}) ({\cal N}_{01}
{\cal N}_{20} + {\cal N}_{00} {\cal N}_{21})
\rbrack
{\cal C}_{11} {\cal C}_{21} 
\nonumber\\
&&+
\lbrack
\beta_7 ({\cal N}_{11} {\cal N}_{20} + {\cal N}_{10}
{\cal N}_{21}) +
\beta_{9} ({\cal N}_{01} {\cal N}_{20} + {\cal N}_{00}
{\cal N}_{21}) 
+ 2 \beta_{16} {\cal N}_{10} {\cal N}_{11} 
\nonumber\\ &&
+
2 \beta_{21} {\cal N}_{00} {\cal N}_{01} 
+ \beta_{34} ({\cal N}_{01} {\cal N}_{10} + {\cal N}_{00}
{\cal N}_{11}) +
4 \beta_{41}
{\cal N}_{20} {\cal N}_{21}
\rbrack
{\cal C}^2_{21}  \qquad \qquad \;\;\;\;\;\;\;\;\;
\;\;\;\;\;\;\;\;\;\;\;\;\;\;\;\;\;\;\;\;\;\;\;\;\;\;\;
\;\;\;\;\;\;\;\;\;\;\;\;\;\;\;\;\
(B.2d) \cr
\nonumber\\
\sigma_{11} &= &(\beta_1 + \beta_2) {\cal N}_{01} {\cal N}_{11}^3 
+ (\beta_3 + \beta_4)  {\cal N}_{21} {\cal N}^3_{11} 
+ \beta_5 {\cal N}_{11} {\cal N}^3_{01} 
+ \beta_6 {\cal N}_{21} {\cal N}^3_{01} 
+ \beta_7 + \beta_8) {\cal N}_{11}
{\cal N}^3_{21} 
+ (\beta_9 + \beta_{10}) {\cal N}_{01} {\cal N}^3_{21} 
\nonumber\\
&&+ (\beta_{11} + \beta_{12} + \beta_{13} + \beta_{14})
{\cal N}^2_{01} {\cal N}^2_{11} 
+ (\beta_{15} + \beta_{16} + \beta_{17} + \beta_{18} + \beta_{19})
{\cal N}^2_{11} {\cal N}^2_{21} 
+ (\beta_{20} + \beta_{21} + \beta_{22} + \beta_{23})
{\cal N}^2_{01} {\cal N}^2_{21} 
\nonumber\\
&&+ (\beta_{24} + \beta_{25} + \beta_{26} + \beta_{27} + \beta_{28})
{\cal N}_{01} {\cal N}_{21} {\cal N}^2_{11} 
+ (\beta_{29} + \beta_{30} + \beta_{31} + \beta_{32})
{\cal N}_{11} {\cal N}_{21} {\cal N}^2_{01} 
\nonumber\\ &&
+ (\beta_{33} + \beta_{34} + \beta_{35} + \beta_{36} + \beta_{37})
{\cal N}_{01} {\cal N}_{11} {\cal N}^2_{21} 
+ \beta_{38} {\cal N}^4_{01} 
+ (\beta_{39} + \beta_{40}) {\cal N}^4_{11} 
+ (\beta_{41} + \beta_{42}) {\cal N}^4_{21} \qquad 
\;\;\;\;\;\;\;\;\;\;\;\;\;\;\;
(B.2e)\cr
\nonumber\\
\sigma_{12} &= &(\beta_1 + \beta_2)
({\cal N}_{02} {\cal N}_{11}^3 + 3 {\cal N}_{01} {\cal N}_{12}
{\cal N}^2_{11}) 
+ (\beta_3 + \beta_4) ({\cal N}_{22} {\cal N}^3_{11}
+ 3 {\cal N}_{21} {\cal N}_{12} {\cal N}_{11})
+ \beta_5 ({\cal N}_{12} {\cal N}^3_{01} + 3 {\cal N}_{11}
{\cal N}_{02} {\cal N}^2_{01})
\nonumber\\
&&+ \beta_6 ({\cal N}_{22} {\cal N}^3_{01}  + 3 {\cal N}_{21}
{\cal N}_{02} {\cal N}^2_{01})
+ (\beta_7 + \beta_8) ({\cal N}_{12}
{\cal N}^3_{21} + 3 {\cal N}_{11} {\cal N}_{22} {\cal N}^2_{21})
+ (\beta_9 + \beta_{10}) ({\cal N}_{02} {\cal N}^3_{21} +
3 {\cal N}_{21} {\cal N}_{22} {\cal N}^2_{21}) 
\nonumber\\
&&+ (\beta_{11} + \beta_{12} + \beta_{13} + \beta_{14})
({\cal N}_{02} {\cal N}_{01} {\cal N}^2_{11} + {\cal N}_{12}
{\cal N}_{11} {\cal N}^2_{01}) 
\nonumber\\ &&
+ 2 (\beta_{15} + \beta_{16} + \beta_{17} + \beta_{18} + \beta_{19})
({\cal N}_{12} {\cal N}_{11} {\cal N}^2_{21} + {\cal N}_{22}
{\cal N}_{21} {\cal N}^2_{11})
\nonumber\\ &&
+ 2 (\beta_{20} + \beta_{21} + \beta_{22} + \beta_{23})
({\cal N}_{02} {\cal N}_{01} {\cal N}^2_{21} + {\cal N}_{22}
{\cal N}_{21} {\cal N}^2_{01})
\nonumber\\
&&+ (\beta_{24} + \beta_{25} + \beta_{26} + \beta_{27} + \beta_{28})
({\cal N}_{02} {\cal N}_{21} {\cal N}^2_{11} + {\cal N}_{01}
{\cal N}_{22} {\cal N}^2_{11} + 2 {\cal N}_{01} {\cal N}_{21}
{\cal N}_{12} {\cal N}_{11}) 
\nonumber\\ &&
+ (\beta_{29} + \beta_{30} + \beta_{31} + \beta_{32})
({\cal N}_{12} {\cal N}_{21} {\cal N}^2_{01} + {\cal N}_{11}
{\cal N}_{22} {\cal N}^2_{01} + 2 {\cal N}_{11} {\cal N}_{21}
{\cal N}_{02} {\cal N}_{01})
\nonumber\\ &&
+ (\beta_{33} + \beta_{34} + \beta_{35} + \beta_{36} + \beta_{37})
({\cal N}_{02} {\cal N}_{11} {\cal N}^2_{21} + {\cal N}_{01}
{\cal N}_{12} {\cal N}^2_{21} + 2 {\cal N}_{01} {\cal N}_{11}
{\cal N}_{22} {\cal N}_{21})
\nonumber\\ &&
+ 4 \beta_{38} {\cal N}_{02} {\cal N}^3_{01} 
+ 4 (\beta_{39} + \beta_{40}) {\cal N}_{12} {\cal N}^3_{11} 
+ 4 (\beta_{41} + \beta_{42}) {\cal N}_{22} {\cal N}^3_{21}.
\qquad \qquad \;\;\;\;\;\;\;\;\;\;\;\;\;\;\;\;\;\;\;\;\;\;\;\;\;\;\;
\;\;\;\;\;\;\;\;\;\;\;\;\;\;\;\;\;\;\;\;\;\;\;\;\;
(B.2f)
\eea*
$\sigma_7$ can be obtained from the formula for $\sigma_1$ and
$\sigma_8$ from the formula for $\sigma_2$ by the substitution
${\cal N}_{i0} \to {\cal N}_{i1}$. $\sigma_5$ and $\sigma_6$ are
constructed from $\sigma_3$ and $\sigma_4$ respectively replacing
${\cal N}_{i1} \to {\cal N}_{i2}$, to find $\sigma_9$ and $\sigma_{10}$
one has to replace ${\cal N}_{i0} \to {\cal N}_{i2}$ in $\sigma_3$
and $\sigma_4$.\par
The ${\cal C}_{ij}$ and ${\cal N}_{ij}$ are again the elements of the
mixing matrix (2.2) and (2.3) and the $\beta_i$ are the couplings
of the unmixed gauge bosons [we use the notation: 
$t \equiv [48 (4 \pi)^2]^{-1} \ \ln(\Lambda/M_W)$]:
\bea*
\beta_1 &=& {1 \over 2} t (\lambda^8 - 2 \lambda^6 - 2 \lambda^4 +
10 \lambda^2 - 7)\cr
\nonumber\\
\beta_2 &=& {1 \over 4} t (\lambda^8 + 4 \lambda^6 - 26 \lambda^4 +
52 \lambda^2 - 31)\cr
\nonumber\\
\beta_3 &=& - {1 \over 2} t (2 \lambda^8 - 5 \lambda^6 + \lambda^4 +
9 \lambda^2 + 1)\cr
\nonumber\\
\beta_4 &=& - {1 \over 2} t (\lambda^8 - \lambda^6 - 7 \lambda^4 +
21 \lambda^2 + 2)\cr
\nonumber\\
\beta_5 &=& {3 \over 4} t (\lambda^8  - 10 \lambda^4 +
24 \lambda^2 - 15)\cr
\nonumber\\
\beta_6 &=& - {3 \over 2} t (\lambda^8  - 2 \lambda^6 - 2 \lambda^4 +
10 \lambda^2 + 1)\cr
\nonumber\\
\beta_7 &=& - 2 t (2 \lambda^8 - \lambda^6 + 1) \ , \qquad
\beta_8 = - 2 t (\lambda^8 + \lambda^6 + 2)\cr
\nonumber\\
\beta_9 &=& \beta_7 \ , \qquad
\beta_{10}= \beta_8\cr
\nonumber\\
\beta_{11} &=& {1 \over 8} t (\lambda^8 - 8 \lambda^6 + 30 \lambda^4 -
40 \lambda^2 + 17)\cr
\nonumber\\
\beta_{12} &=& {1 \over 4} t (\lambda^8 + 4 \lambda^6 + 6 \lambda^4 -
28 \lambda^2 + 17)\cr
\nonumber\\
\beta_{13} &=& {1 \over 4} t (\lambda^8 - 10 \lambda^6 + 34 \lambda^4 -
42 \lambda^2 + 17)\cr
\nonumber\\
\beta_{14} &=& {1 \over 2} t (\lambda^8 + 8 \lambda^6 - 2 \lambda^4 -
24 \lambda^2 + 17)\cr
\nonumber\\
\beta_{15} &=& {1 \over 2} t (\lambda^8 - 3 \lambda^6 + 4 \lambda^4 +
\lambda^2 + 1)\cr
\nonumber\\
\beta_{16} &=& t (\lambda^8 +  \lambda^4 + 7 \lambda^2 + 1)\cr
\nonumber\\
\beta_{17} &=& {1 \over 2} t (2 \lambda^8 - 7 \lambda^6 + 10 \lambda^4 + \lambda^2 + 2)\cr
\nonumber\\
\beta_{18} &=& {1 \over 2} t (2 \lambda^8 + 5 \lambda^6 - 14 \lambda^4 + 13 \lambda^2 + 2)\cr
\nonumber\\
\beta_{19} &=& {1 \over 2} t (2 \lambda^8 - \lambda^6 - 8 \lambda^4 +
\lambda^2 + 2)\cr
\nonumber\\
\beta_{20} &=& \beta_{15} \ , \quad
\beta_{21} = \beta_{16} \ , \quad
\beta_{22} = \beta_{17} \ , \quad
\beta_{23} = \beta_{18}\cr
\eea*
\bea*
\beta_{24} &=& - {1 \over 2} t (\lambda^8 - 3 \lambda^6 + 9 \lambda^4 -
9 \lambda^2 + 2)\cr
\nonumber\\
\beta_{25} &=& - {1 \over 2} t (2 \lambda^8 + 3 \lambda^6 + 3 \lambda^4 - 9 \lambda^2 + 1)\cr
\nonumber\\
\beta_{26} &=& - t (\lambda^8 + 8 \lambda^6 - 13 \lambda^4 +
2 \lambda^2 + 2)\cr
\nonumber\\
\beta_{27} &=& - {1 \over 2} t (2 \lambda^8 + \lambda^6 - 5 \lambda^4 +
\lambda^2 + 1)\cr
\nonumber\\
\beta_{28} &=& - {1 \over 2} t (2 \lambda^8 - 11 \lambda^6 + 25 \lambda^4 - 17 \lambda^2 + 1)\cr
\nonumber\\
\beta_{29} &=& \beta_{24} \ , \quad
\beta_{30} = \beta_{25} \cr 
\nonumber\\
\beta_{31} &=& - {1 \over 2} t (4 \lambda^8 + 17 \lambda^6 - 31 \lambda^4 + 5 \lambda^2 + 5)\cr
\nonumber\\
\beta_{32} &= & \beta_{28}\cr
\nonumber\\
\beta_{33} &=& t (\lambda^8 + 7 \lambda^6 - 8 \lambda^4
- \lambda^2 + 1)\cr
\nonumber\\
\beta_{34} &=& 2 t (\lambda^8 - 2 \lambda^6 +  \lambda^4
- 7 \lambda^2 + 1)\cr
\nonumber\\
\beta_{35} &=& t (2 \lambda^8 + 3 \lambda^6 +  6 \lambda^4
- 13 \lambda^2 + 2)\cr
\nonumber\\
\beta_{36} &=& t (2 \lambda^8 - 3 \lambda^6 + 12 \lambda^4
- \lambda^2 + 2)\cr
\nonumber\\
\beta_{37} &=& t (2 \lambda^8 + 3 \lambda^6 - 6 \lambda^4
- \lambda^2 + 2)\cr
\nonumber\\
\beta_{38} &=& {3 \over {16}} t (\lambda^8 - 4 \lambda^6 + 6 \lambda^4
- 4 \lambda^2 + 17)\cr
\nonumber\\
\beta_{39} &=& {1 \over 8} t (\lambda^8 - 4 \lambda^6 + 2 \lambda^4
+ 4 \lambda^2 - 3) + {1 \over {4g^2}}\cr
\nonumber\\
\beta_{40} &=& {1 \over {16}} t (\lambda^8 - 4 \lambda^6 + 14 \lambda^4
- 20 \lambda^2 + 57) - {1 \over {4g^2}}\cr
\nonumber\\
\beta_{41} &=& 2 t (\lambda^8 - \lambda^4)
+ {1 \over {g^{\prime\prime 2}}}\cr
\nonumber\\
\beta_{42} &=&  t (\lambda^8 + 2 \lambda^4 + 3)
- {1 \over {g^{\prime\prime 2}}}
\qquad \;\;\;\;\;\;\;\;\;\;\;\;\;\;\;\;\;\;\;\;\;\;\;\;\;\;\;\;\;\;\;\;\;\;
\;\;\;\;\;\;\;\;\;\;\;\;\;\;\;\;\;\;\;\;\;\;\;\;\;\; (B.3)
\eea*

\newpage

\begin{figure}[htb]
\centering\includegraphics[width=120mm,height=240mm]{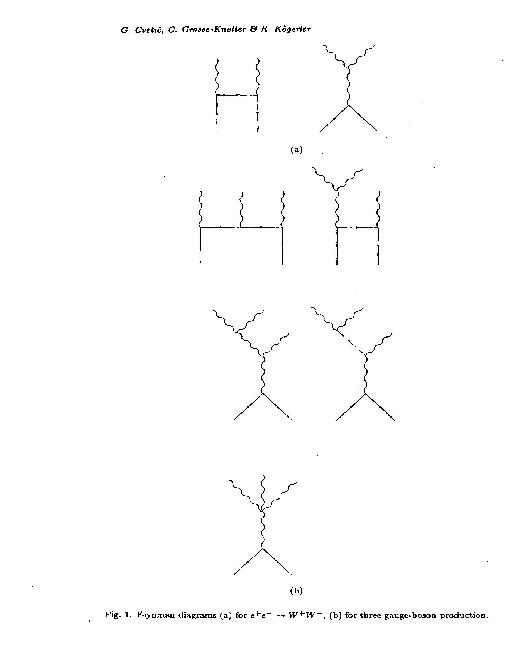}
\vspace{-0.4cm}
\caption{Feynman diagrams: (a) for $e^+e^- \to W^+W^-$; 
(b) for three gauge boson production.}
\label{fig1}
\end{figure}

\begin{figure}[htb]
\centering\includegraphics[width=120mm,height=240mm]{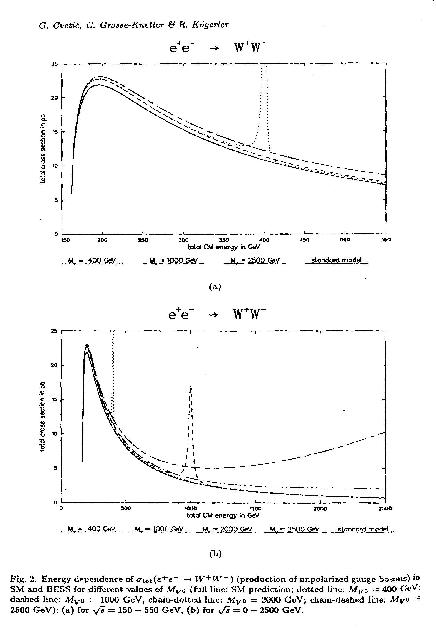}
\vspace{-0.4cm}
\caption{Energy dependence of $\sigma_{tot} (e^+e^- \to W^+W^-)$ (production
of unpolarized\break gauge bosons) in SM and BESS for different values of 
$M_{V^0}$
(full line: SM prediction; dotted line: $M_{V^0} = 400 \ {\rm GeV}$;
dashed line: $M_{V^0} = 1000 \ {\rm GeV}$; chaindotted line:
$M_{V^0} = 2000 \ {\rm GeV}$;
chaindashed line: $M_{V^0} = 2500 \ {\rm GeV}$); (a) for $\sqrt s = 150 - 550 \ {\rm GeV}$;
(b) for $\sqrt s = 0 - 2500 \ {\rm GeV}$.}
\label{fig2}
\end{figure}

\begin{figure}[htb]
\centering\includegraphics[width=120mm,height=240mm]{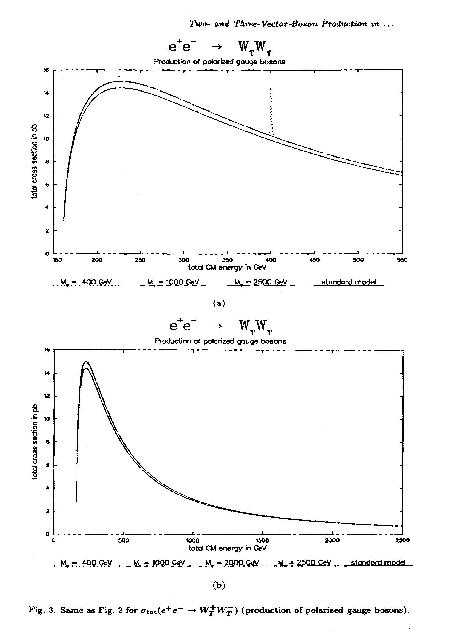}
\vspace{-0.4cm}
\caption{Same as Fig. 2 for $\sigma_{tot} (e^+e^- \to W^+_T W^-_T)$
(production of polarized gauge bosons).}
\label{fig3}
\end{figure}

\begin{figure}[htb]
\centering\includegraphics[width=120mm,height=240mm]{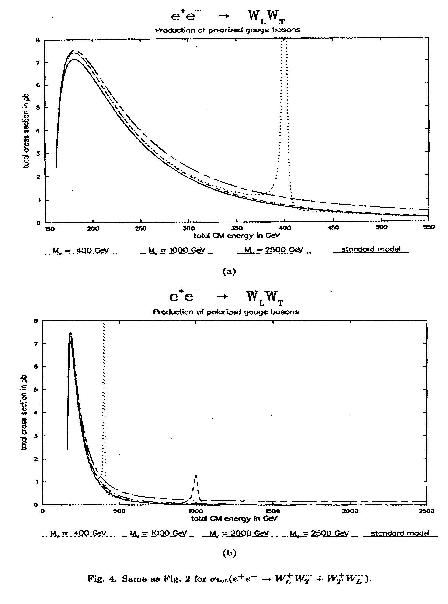}
\vspace{-0.4cm}
\caption{Same as Fig. 2 for $\sigma_{tot}
(e^+e^- \to W^+_L W^-_T + W^+_T W^-_L)$.}
\label{fig4}
\end{figure}

\begin{figure}[htb]
\centering\includegraphics[width=120mm,height=240mm]{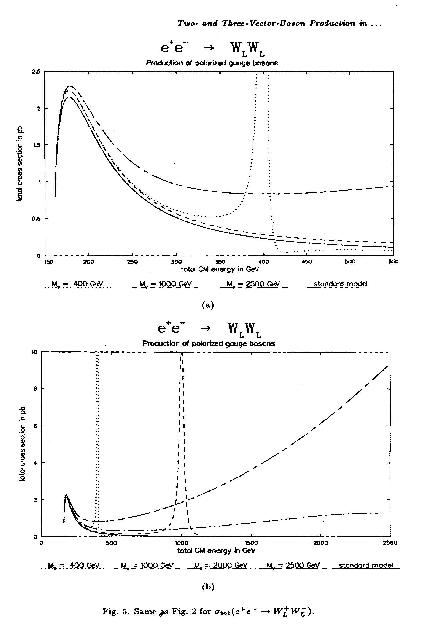}
\vspace{-0.4cm}
\caption{Same as Fig. 2 for $\sigma_{tot}
(e^+e^- \to W^+_L W^-_L)$.}
\label{fig5}
\end{figure}

\begin{figure}[htb]
\centering\includegraphics[width=120mm,height=240mm]{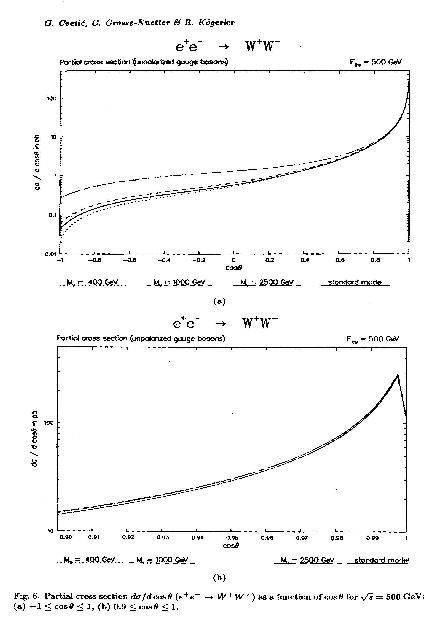}
\vspace{-0.4cm}
\caption{Partial cross section $d \sigma/d \cos \theta (e^+e^- \to W^+W^-)$
as a function of $\cos \theta$ for $\sqrt s = 500 \ {\rm GeV}$: (a) $- 1 \le \cos \theta \le 1$;
(b) $0.9 \le \cos \theta \le 1$.}
\label{fig6}
\end{figure}

\begin{figure}[htb]
\centering\includegraphics[width=120mm,height=240mm]{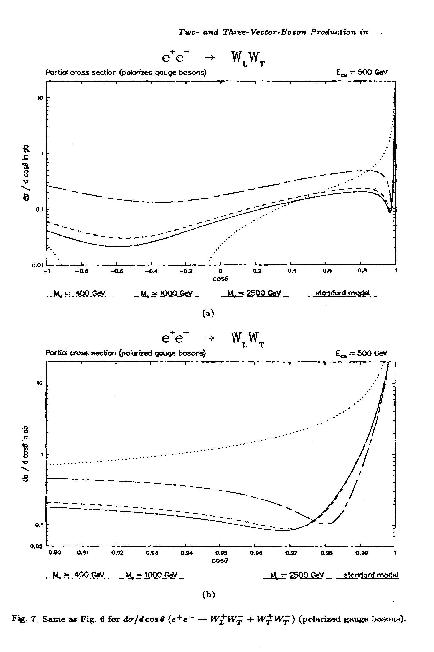}
\vspace{-0.4cm}
\caption{Same as Fig. 6 for $d \sigma/d \cos \theta (e^+e^- \to W^+_L W^-_T +
W^+_T W^-_T)$ (polarized gauge bosons).}
\label{fig7}
\end{figure}

\begin{figure}[htb]
\centering\includegraphics[width=120mm,height=240mm]{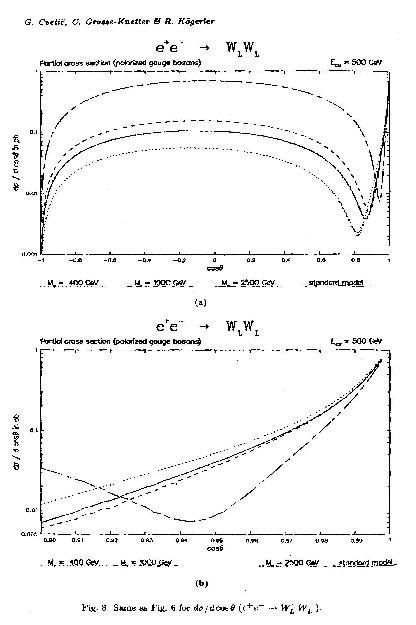}
\vspace{-0.4cm}
\caption{Same as Fig. 6 for $d \sigma/d \cos \theta (e^+e^- \to W^+_L W^-_L)$.}
\label{fig8}
\end{figure}

\begin{figure}[htb]
\centering\includegraphics[width=120mm,height=240mm]{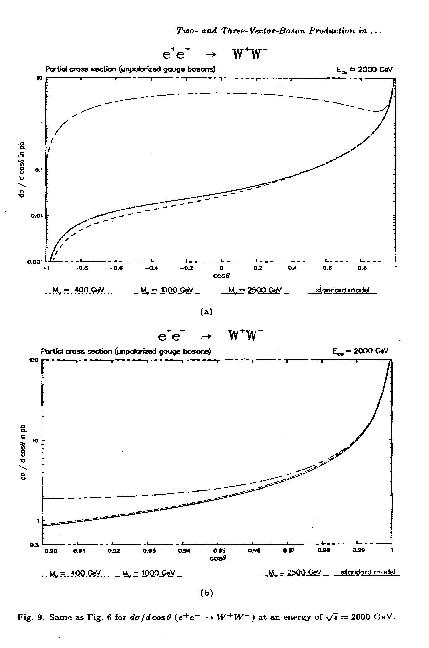}
\vspace{-0.4cm}
\caption{Same as Fig. 6 for $d \sigma/d \cos \theta (e^+e^- \to W^+ W^-)$
at an energy of $\sqrt s = 2000 \ {\rm GeV}$.}
\label{fig9}
\end{figure}

\begin{figure}[htb]
\centering\includegraphics[width=120mm,height=240mm]{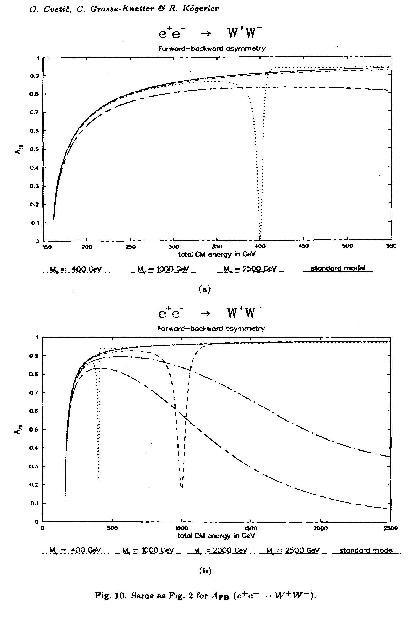}
\vspace{-0.4cm}
\caption{Same as Fig. 2 for $A_{FB} (e^+e^- \to W^+W^-)$.}
\label{fig10}
\end{figure}

\begin{figure}[htb]
\centering\includegraphics[width=120mm,height=240mm]{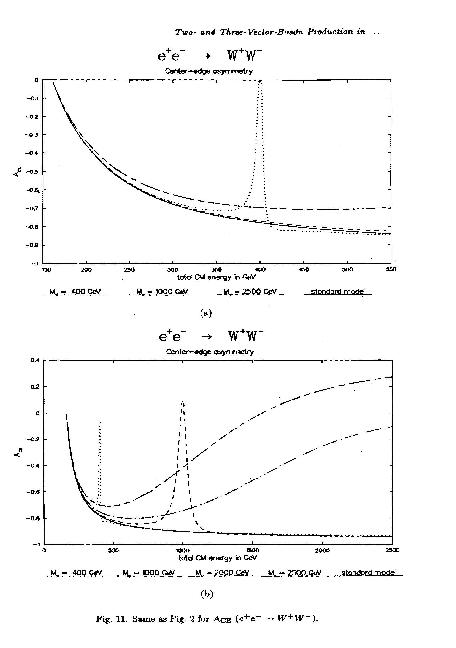}
\vspace{-0.4cm}
\caption{Same as Fig. 2 for $A_{CE} (e^+e^- \to W^+W^-)$.}
\label{fig11}
\end{figure}

\begin{figure}[htb]
\centering\includegraphics[width=120mm,height=240mm]{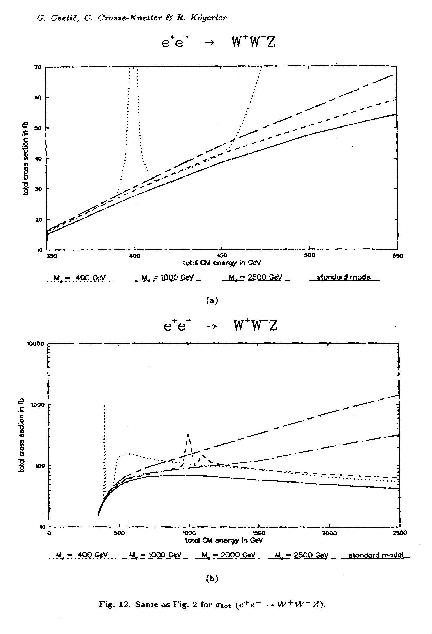}
\vspace{-0.4cm}
\caption{Same as Fig. 2 for $\sigma_{tot} (e^+e^- \to W^+W^-Z)$.}
\label{fig12}
\end{figure}

\begin{figure}[htb]
\centering\includegraphics[width=120mm,height=240mm]{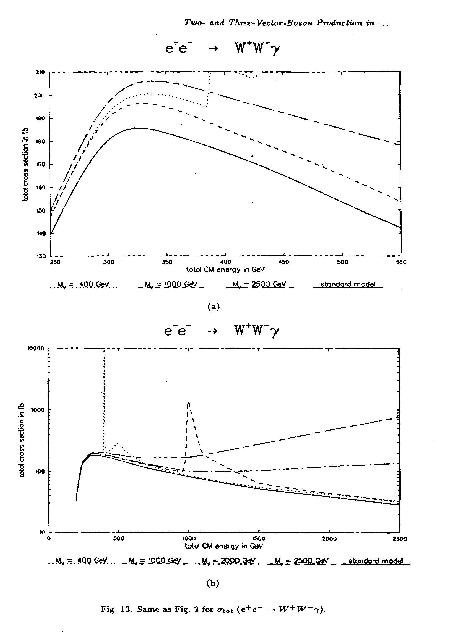}
\vspace{-0.4cm}
\caption{Same as Fig. 2 for $\sigma_{tot} (e^+e^- \to W^+W^-\gamma)$.}
\label{fig13}
\end{figure}

\begin{figure}[htb]
\centering\includegraphics[width=120mm,height=240mm]{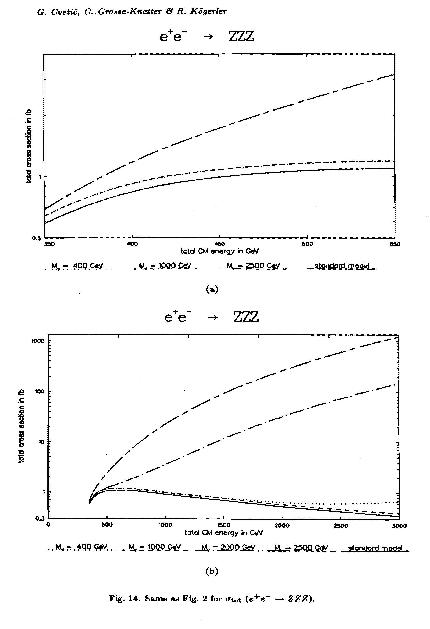}
\vspace{-0.4cm}
\caption{Same as Fig. 2 for $\sigma_{tot} (e^+e^- \to ZZZ)$.}
\label{fig14}
\end{figure}

\begin{figure}[htb]
\centering\includegraphics[width=120mm,height=240mm]{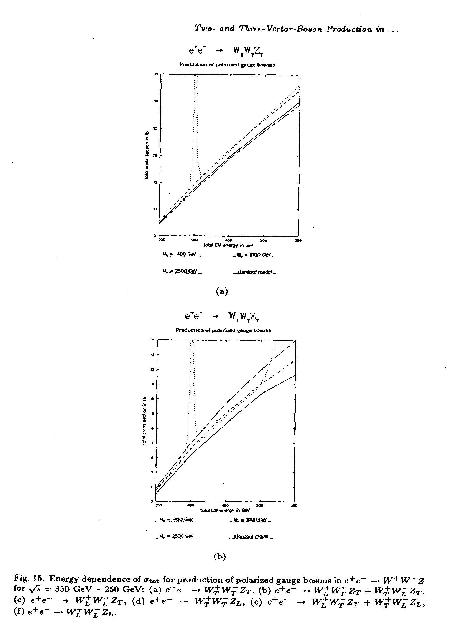}
\vspace{-0.4cm}
\caption{Energy dependence of $\sigma_{tot}$ for production of polarized
gauge bosons in
$e^+e^- \to W^+W^- Z$ for $\sqrt s = 350 - 250 \ {\rm GeV}$:
(a) $e^+e^- \to W^+_T W^-_T Z^{~}_T$; (b) $e^+e^- \to W^+_L W^-_T Z^{~}_T
+ W^+_T W^-_L Z^{~}_T$;
(c) $e^+e^- \to W^+_L W^-_L Z^{~}_T$;
(d) $e^+e^- \to W^+_T W^-_T Z^{~}_L$;
(e) $e^+e^- \to W^+_L W^-_T Z^{~}_L
+ W^+_T W^-_L Z^{~}_L$;
(f) $e^+e^- \to W^+_L W^-_L Z^{~}_L$.}
\label{fig15ab}
\end{figure}

\begin{figure*}[htb]
\centering\includegraphics[width=120mm,height=240mm]{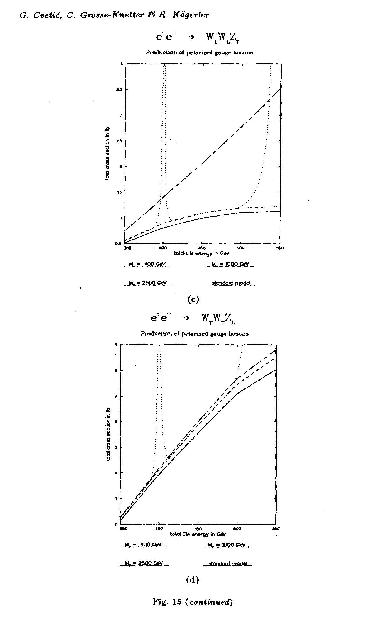}
\caption{(Fig.~15 continued: Figs. 15c,d)}
\label{fig15cd}
\end{figure*}

\begin{figure*}[htb]
\centering\includegraphics[width=120mm,height=240mm]{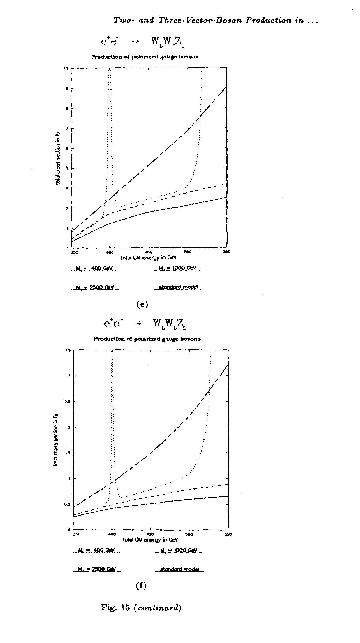}
\caption{(Fig.~15 continued: Figs. 15e,f)}
\label{fig15ef}
\end{figure*}

\begin{figure}[htb]
\centering\includegraphics[width=120mm,height=240mm]{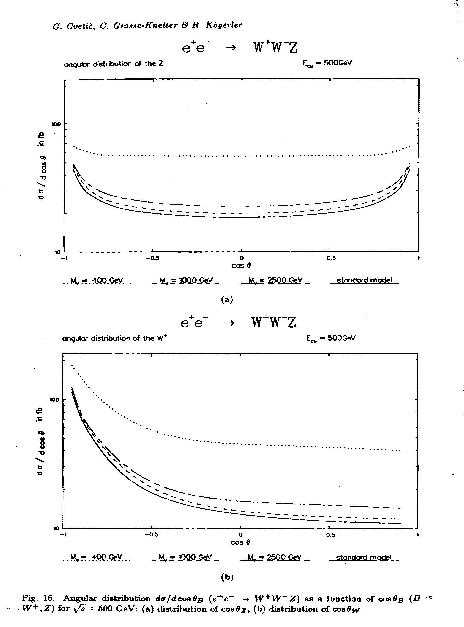}
\vspace{-0.4cm}
\caption{Fig.~16: Angular distribution $d \sigma/d \cos \theta_B (e^+e^- \to W^+W^-Z)$
as a function of $\cos \theta_B\break
(B = W^+, Z)$ for $\sqrt s = 500 \ {\rm GeV}$:
(a) distribution of $\cos \theta_Z$;
(b) distribution of $\cos \theta_W$.}
\label{fig16}
\end{figure}

\begin{figure}[htb]
\centering\includegraphics[width=120mm,height=240mm]{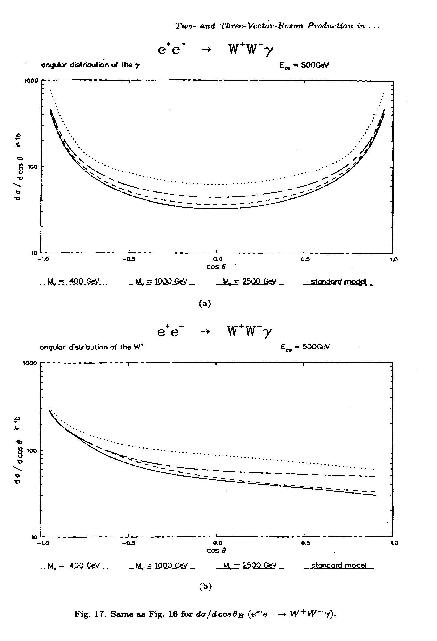}
\vspace{-0.4cm}
\caption{Fig.~17: same as Fig. 16 for $d \sigma/d \cos \theta_B
(e^+e^- \to W^+W^-\gamma)$.}
\label{fig17}
\end{figure}

\begin{figure}[htb]
\centering\includegraphics[width=120mm,height=240mm]{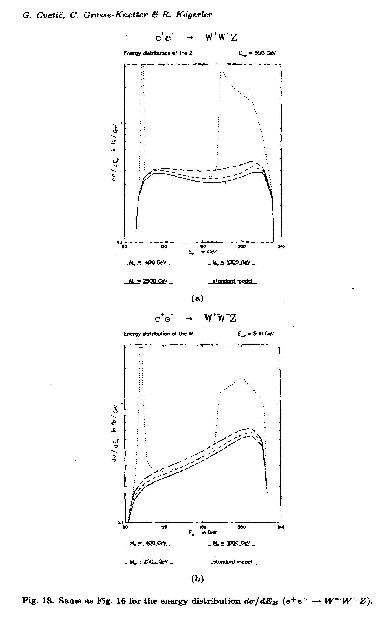}
\vspace{-0.4cm}
\caption{Fig.~18: same as Fig. 16 for the energy distribution $d \sigma/dE_B
(e^+e^- \to W^+W^- Z)$.}
\label{fig18}
\end{figure}

\begin{figure}[htb]
\centering\includegraphics[width=120mm,height=240mm]{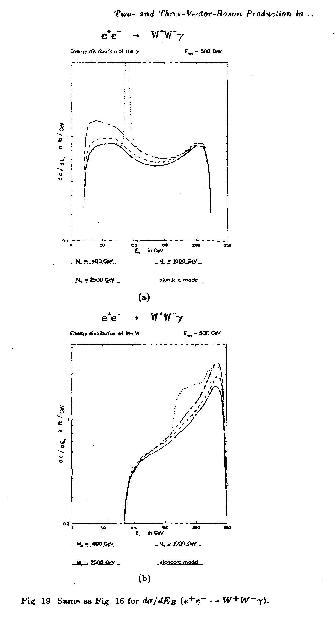}
\vspace{-0.4cm}
\caption{Fig.~19: same as Fig. 16 for $d \sigma/dE_B
(e^+e^- \to W^+W^- \gamma)$.}
\label{fig19}
\end{figure}

\begin{figure}[htb]
\centering\includegraphics[width=120mm,height=240mm]{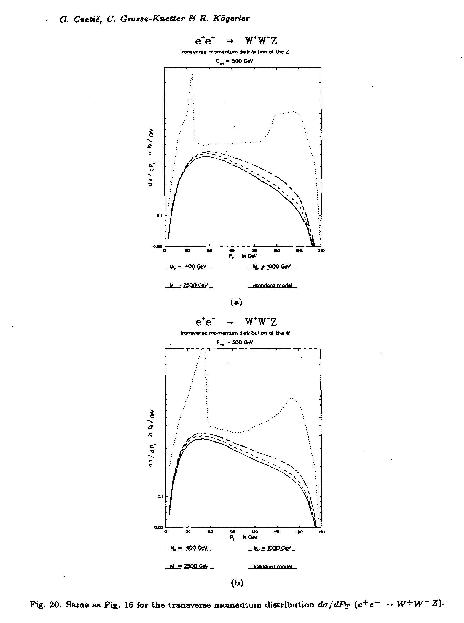}
\vspace{-0.4cm}
\caption{Fig.~20: same as Fig. 16 for the transverse momentum distribution $d \sigma/dP_T$
($e^+e^-$\break $\to W^+W^- Z$).}
\label{fig20}
\end{figure}

\begin{figure}[htb]
\centering\includegraphics[width=120mm,height=240mm]{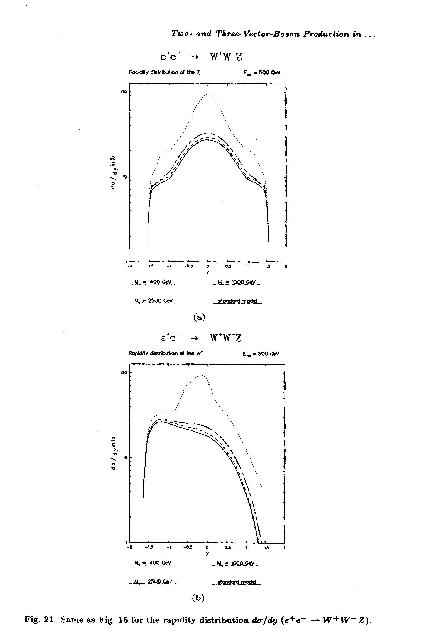}
\vspace{-0.4cm}
\caption{Fig.~21: same as Fig. 16 for the rapidity distribution $d \sigma/dy
(e^+e^- \to W^+W^- Z)$.}
\label{fig21}
\end{figure}


\begin{thebibliography}{99}


\bibitem{Proc1}
See  ``'$e^+ e^-$ Collisions at 500 GeV' - The Physical Impact,'' 
published in DESY-Report-92-123, ed. P.~Zerwas, DESY, Hamburg, Germany, 1992;
Proceedings of the ``Workshop on Physics and Experiments with Linear
Colliders'', ed. M.~Nordberg, Saariselk\"a, Finnland, 1991.


\bibitem{Prep2}
For cubic self-interactions a fairly model-dependent
analysis based on symmetry considerations has been performed in:
G.~Gounaris, J.~L.~Kneur, J.~Layssac, G.~Moultaka, F.~M.~Renard,
D.~Schildknecht (Bielefeld-Montpellier-Thessaloniki-Collaboration),
in Ref.~\cite{Proc1}, p.735 (and: Bielefeld preprint BI-TP 91/40 (1991));
  M.~S.~Bilenky, J.~L.~Kneur, F.~M.~Renard and D.~Schildknecht,
  ``Trilinear couplings among the electroweak vector bosons and their determination at LEP-200,''
  Nucl.\ Phys.\ B\ {\bf 409}, 22  (1993).

\bibitem{UGVA-DPT-1986-01-492} 
  R.~Casalbuoni, S.~De Curtis, D.~Dominici and R.~Gatto,
  ``Physical implications of possible $J=1$ bound states from strong Higgs,''
  Nucl.\ Phys.\ B\ {\bf 282}, 235  (1987);
  ``Effective weak interaction theory with possible new vector resonance from a strong Higgs sector,''
  Phys.\ Lett.\ B\ {\bf 155}, 95  (1985).


\bibitem{BI-TP-88/32} 
  G.~Cveti\v{c} and R.~K\"ogerler,
  ``Fermionic couplings in an electroweak theory with nonlinear spontaneous symmetry breaking,''
  Nucl.\ Phys.\ B\ {\bf 328}, 342  (1989);
  ``Induced fermionic interactions in theories with general nonlinearly realized SSB,''
  Z.\ Phys.\ C\ {\bf 48}, 109  (1990).

\bibitem{BI-TP-90-43} 
  G.~Cveti\v{c} and R.~K\"ogerler,
  ``Complete determination of gauge boson selfinteractions in the BESS model,''
  Nucl.\ Phys.\ B\ {\bf 363}, 401  (1991).

\bibitem{it6}
G. Cveti\v c, C. Grosse-Knetter and R. K\"ogerler,
``Vector boson selfinteractions and vector boson production within the BESS model,'' 
in Ref.~\cite{Proc1}, p.~775
(and Bielefeld preprint BI-TP 91/37 (1991)).


\bibitem{hep-ph/9212291} 
R.~B\"onisch, C.~Grosse-Knetter and R.~K\"ogerler,
``Bounds on BESS model parameters from vector boson production in e+ e- collisions,''
Z.\ Phys.\ C\ {\bf 59}, 109  (1993)
[hep-ph/9212291].


\bibitem{104360} 
  A.~Hosoya and K.~Kikkawa,
  ``Quantum theory of collective motions and an application totTheory of extended hadrons,''
Nucl.\ Phys.\ B\ {\bf 101}, 271  (1975);
  J.~Alfaro and P.~H.~Damgaard,
  ``Field transformations, collective coordinates and BRST invariance,''
  Annals Phys.\ \ {\bf 202}, 398  (1990).

\bibitem{879856} 
  T.~Kunimasa and T.~Goto,
  ``Generalization of the Stueckelberg formalism to the massive Yang-Mills field,''
  Prog.\ Theor.\ Phys.\ \ {\bf 37}, 452  (1967);
  T.~Sonoda and S.~Y.~Tsai,
  ``The generalized Stuckelberg formalism and the Glashow-Weinberg-Aalam electroweak model,''
  Prog.\ Theor.\ Phys.\ \ {\bf 71}, 878  (1984).

\bibitem{YTP-80-01} 
T.~Appelquist and C.~W.~Bernard,
  ``Strongly interacting Higgs bosons,''
  Phys.\ Rev.\ D\ {\bf 22}, 200  (1980);
  A.~C.~Longhitano,
  ``Low-energy impact of a heavy Higgs boson sector,''
  Nucl.\ Phys.\ B\ {\bf 188}, 118  (1981);
  J.~van der Bij and M.~J.~G.~Veltman,
  ``Two loop large Higgs mass correction to the $\rho$ parameter,''
  Nucl.\ Phys.\ B\ {\bf 231}, 205  (1984).


\bibitem{RRK 84-22} 
M.~Bando, T.~Kugo, S.~Uehara, K.~Yamawaki and T.~Yanagida,
  ``Is $\rho$ meson a dynamical gauge boson of hidden local symmetry?,''
  Phys.\ Rev.\ Lett.\ \ {\bf 54}, 1215  (1985);
see also:
  A.~P.~Balachandran, A.~Stern and C.~G.~Trahern,
  ``Nonlinear models as gauge theories,''
  Phys.\ Rev.\ D\ {\bf 19}, 2416  (1979).

\bibitem{ITEP-62-1978} 
 For two-dimensional theories see:  
V.~L.~Golo and A.~M.~Perelomov,
  ``Solution of the Duality Equations for the two-dimensional SU(N) invariant Chiral Model,''
  Phys.\ Lett.\ B\ {\bf 79}, 112  (1978);
  A.~D'Adda, M.~L\"uscher and P.~Di Vecchia,
  ``A $1/n$-expandable series of Nonlinear Sigma Models with instantons,''
  Nucl.\ Phys.\ B\ {\bf 146}, 63  (1978);
for three-dimensional theories see:
  I.~Y.~.Arefeva and S.~I.~Azakov,
  ``Renormalization and phase transition in the Quantum $C_p^{n-1}$ Model 
($d = 2, 3$),''
  Nucl.\ Phys.\ B\ {\bf 162}, 298  (1980);
for four-dimensional theories see:
  R.~K\"ogerler, W.~Lucha, H.~Neufeld and H.~Stremnitzer,
  ``Dynamical generation of gauge bosons of hidden local symmetries in Nonlinear Sigma Models,''
  Phys.\ Lett.\ B\ {\bf 201}, 335  (1988);
T.~Kugo, Soryushiron Kenkyu 71, E78 (1985);
  T.~Kugo, H.~Terao and S.~Uehara,
  ``Dynamical gauge bosons and hidden local symmetries,''
  Prog. Theor. Phys. Suppl., No.~85, 122-135 (1985).

\bibitem{CERN-TH-4876/87} 
R.~Casalbuoni, P.~Chiappetta, D.~Dominici, F.~Feruglio and R.~Gatto,
  ``High-energy tests for a possible strong sector in the standard model,''
  Nucl.\ Phys.\ B\ {\bf 310}, 181  (1988);
  R.~Casalbuoni, D.~Dominici, F.~Feruglio and R.~Gatto,
  ``Testing the Standard Model in terms of a possible strong scalar sector,''
  Phys.\ Lett.\ B\ {\bf 200}, 495  (1988).

\bibitem{BI-TP-90-40} 
  R.~B\"onisch and R.~K\"ogerler,
  ``Phenomenological parameter analysis of a Higgsless electroweak gauge theory with induced interactions included,''
  Int.\ J.\ Mod.\ Phys.\ A\ {\bf 7}, 5475  (1992).

\bibitem{UT-KOMABA 76-12} 
  K.~-i.~Shizuya,
  ``Renormalization of two-dimensional massive Yang-Mills theory and nonrenormalizability of its four-dimensional version,''
  Nucl.\ Phys.\ B\ {\bf 121}, 125  (1977).

\bibitem{hep-ph/9303201} 
  R.~Casalbuoni, P.~Chiappetta, A.~Deandrea, S.~De Curtis, D.~Dominici and R.~Gatto,
  ``Vector resonances from a strong electroweak sector at linear colliders,''
  Z.\ Phys.\ C\ {\bf 60}, 315  (1993)
  [hep-ph/9303201].

\bibitem{Leenen}
M.~Leenen, talk given at EE500 Workshop (MPI, Munich, Germany, 1992).

\bibitem{MAD/PH/420} 
V.~D.~Barger, T.~Han and R.~J.~N.~Phillips,
  ``$W W Z$, $Z Z Z$ and $W W \gamma$ production at $e^+ e^-$ colliders,''
  Phys.\ Rev.\ D\ {\bf 39}, 146  (1989).

\bibitem{UCD-88-24} 
  A.~Tofighi-Niaki and J.~F.~Gunion,
  ``General amplitudes for three gauge boson production and cross sections and polarization analysis at an $e^+ e^-$ collider,"
  Phys.\ Rev.\ D\ {\bf 39}, 720  (1989).

\bibitem{FMSZ}
 M. Frank, P. M\"attig, R. Settles, W. Zeuner,
in Ref.~\cite{Proc1}, p.~223.

\bibitem{DESY 85/133} 
  K.~Hagiwara and D.~Zeppenfeld,
  ``Helicity amplitudes for heavy lepton production in $e^+ e^-$ annihilation,''
  Nucl.\ Phys.\ B\ {\bf 274}, 1  (1986).

\bibitem{BI-TP-90/14} 
For first (incomplete) calculation see: 
  G.~Cveti\v{c}, R.~K\"ogerler and J.~Trampeti\'c,
  ``Phenomenological implications of an electroweak theory without Higgs,''
  Phys.\ Lett.\ B\ {\bf 248}, 128  (1990).


\end{thebibliography}
\end{document}